\documentclass[12pt]{elsarticle}
\usepackage{amssymb}
\usepackage[svgnames*,table,dvipsnames]{xcolor} 

\journal{Optics and Lasers in Engineering}
\begin{document}
\begin{frontmatter}
\title{Seeding optimization for instantaneous volumetric velocimetry.\\ Application to a jet in crossflow}
\author{Tristan Cambonie$^{*}$}
\author{Jean-Luc Aider}
\address{$^{*}$tristan.cambonie@espci.fr, aider@pmmh.espci.fr,\\Laboratoire PMMH, CNRS UMR7636 / ESPCI / UPMC / UPD , \a'Ecole Sup\a'erieure de Physique et de Chimie Industrielles (ESPCI), 10 rue Vauquelin, 75005 Paris, France}

\begin{abstract}
Every volumetric velocimetry measurements based on tracer (particles, bubbles, etc.) detection can be strongly influenced by the optical screening phenomenon. It has to be taken into account when the the statistical properties associated to the performances of the particle detection and tracking algorithms are significantly affected. It leads to a maximum concentration of particles in the images thus limiting the final spatial resolution of the instantaneous three-dimensional three-components (3D3C) velocity fields. A volumetric velocimetry system based on Defocused Digital Particle Image Velocimetry  (DDPIV), named Volumetric 3-Components Velocimetry (V3V), is used to show that above a critical visual concentration of particles in the images,  the concentration and accuracy of the final instantaneous raw velocity vector field drop. The critical concentration depends on physical parameters as well as on the processing algorithms. Three distinct regimes  are identified. In the first regime, the concentration is well adapted to volumetric velocimetry, the largest concentration being optimal to maximize the number of valid velocity vectors. In the second regime, the performances of the detection and tracking algorithms are no longer optimal. Finally, the third regime is strongly influenced by optical screening. A rigorous methodology is proposed to optimize volumetric measurements taking into account the system specifications (pixel size, focal length, system magnification, etc.). An optimal particle  concentration of 4.5$\times 10^{-2}$~particles.pixel$^{-1}$ (ppp) is found. A relationship between the imaged concentration and the optimal final spatial resolution for fully interpolated  instantaneous velocity field is obtained. The present study can be used as a guideline to achieve the measurement of instantaneous three-dimensional velocity fields with a good spatial resolution.  It is finally successfully applied to a complex 3D flow test case: a jet in cross-flow.
\end{abstract}

\end{frontmatter}

\section{Introduction}
During the last two decades the various methods of particles imaging using a planar laser sheet, like Particle Image Velocimetry (PIV), have proven to be powerful measuring techniques. Measuring instantaneous two components velocity fields in a plane (2D2C) is now very common. The improvements of the different parts of the set-ups (pulsed Lasers, double-frame or time-resolved cameras, computers, etc.) lead to more and more sophisticated measurements: largest measurement areas, access to the three components of the velocity fields in a plane (2D3C) and increasing spatial and temporal resolutions. 
Nowadays, the ultimate objective is to measure instantaneous 3D3C velocity fields (3 velocity Components in a 3D volume) with a good spatial resolution together with a high sampling frequency. 

During the last few years, many innovative techniques have been developed to measure instantaneous 3D3C velocity fields: Holographic PIV (HPIV) (\citet{Meng2004}), Tomographic PIV (TomoPIV) (\citet{Elsinga2006}), or Defocusing Digital PIV (DDPIV) (\citet{willert1992,Pereira2000}). Thanks to the increasing computing power, the volumetric velocimetry should become very popular in the near future. Nevertheless, the experimentalists still have to face new challenges. 

Some of the constraints of the 2D velocimetry techniques no longer exist when measuring the flow velocity in a volume. For example, the careful spatial orientation of the laser sheet together with the fine tuning of its thickness are no longer needed. Unfortunately, adding a third dimension also leads to new issues: ''ghost'' particles (\citet{Graff2008}), calibration of the measurement volume, localization of the particles in the volume, image distortions, magnification changes along the measurement volume depth, perspective errors, post-processing of a very large amount of data, etc. If some solutions were found (telecentric lenses, improvement of the algorithms, etc), the intrinsic problems of 3D measurements still remain and are still a limitation to their development.

Since the whole 3D particles field is projected over the 2D sensor of the camera, the real 3D concentration of particles per unit of volume has to be lowered compared to the concentration of particles used for standard 2D PIV measurements. This is a consequence of the screening phenomenon. If a particle can mask one or more other particles even at low particle concentration, it has to be taken into account only when it becomes statistically relevant and can then lead to false estimations of the 3D particles field. In the following, the screening phenomenon will only be considered from a statistical point of view. Working with a low or medium particles concentration (very low compared to the concentration used for standard 2D2C PIV measurements) is a solution to avoid the screening phenomenon, but it leads to instantaneous velocity fields with many holes and a poor spatial resolution. 

A good spatial resolution can be obtained only when considering  time-averaged velocity fields. It is relevant only if the flow is stationary or if the flow is periodic. In this case, phase averaging (\citet{Sharp2010}) or repetition of the same experiment for a highly reproducible flow (\citet{Troolin2010,Kim2010}) can be a good solution to obtain a high spatial resolution. 
Otherwise, a compromise has to be found between the spatial resolution, the particle concentration and the thickness of the illumination volume.

The objective of the present work is to optimize the seeding of the flow to measure \emph{instantaneous}  3D3C velocity fields with a good spatial resolution for a given illumination. This study has been carried out using a DDPIV-3DPTV system similar to the one used by \citet{Pereira2006}, designed by TSI and named  Volumetric 3-Components Velocimetry (V3V).

To evaluate the efficiency of the processing steps one can define the parameter $R_{Eff Proc}$ which is the ratio of the final number of raw velocity vectors over the number of 2D particles detected within the images.
 $R_{Eff Proc} =$ 0 corresponds to the case where no valid velocity vectors are found while $R_{Eff Proc} =$ 1 corresponds to the ideal situation where every 2D particles lead to a valid velocity vector.  In practice,  $R_{Eff Proc}$ ranges in the literature between $1/4$ (\citet{Sharp2010}) and $1/5$ (\citet{Wolf2011}). It means that, in the best case, only a fourth of the detected particles is successfully tracked and provides raw velocity vectors. This ratio can also be interpreted as the link between the effective concentration in detected particles and the final concentration in raw velocity vectors. In the limit of the very small concentrations (where the inter-particle distance is large compared to the maximum displacement of a particle between two time steps), the processing efficiency improves to reach ultimately $R_{Eff Proc} = 1$. The objective of this work is to obtain $R_{Eff Proc} > $ $1/4$ for large particle concentrations.

In the first two sections, the experimental setup and the measurement method are introduced. The characteristics of the hydrodynamic channel are presented while the choices of the particles and associated uncertainties are detailed. In the next section, the influence of the particle concentration is studied and is related to the final size of interpolation. Finally, in the last section, we validate the methodology using a jet-in-crossflow test case.

\section{Experimental method}
\subsection{Experimental setup and parameters}
\begin{figure}[!htb]
\begin{center}
\includegraphics[width=0.8\textwidth]{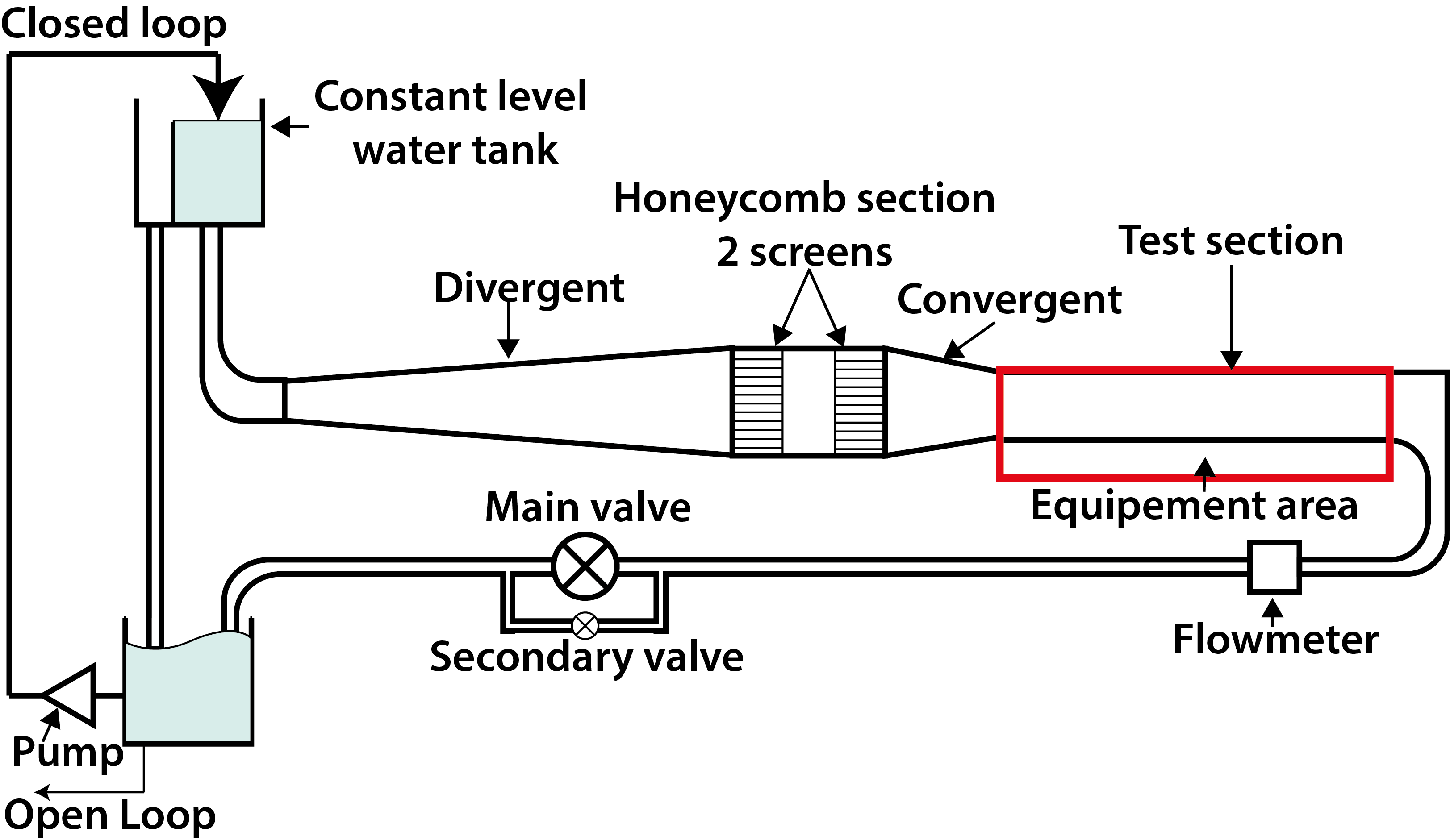}
\end{center}
\caption{Schematic description of the hydrodynamic channel.}\label{Fig1}
\end{figure}

A low-speed closed-loop water tunnel was used for this investigation (Fig. \ref{Fig1}).  The flow is driven by gravity  using  a constant-level water tank. A divergent part, two honeycombs and a convergent section reduce the free-stream turbulence and suppress undesired large structures. The  free-stream velocities $U_{\infty}$  ranges between $0$ and 22 cm.s$^{-1}$.  The turbulence level $T_{turb} \approx 0.2 \%$ is computed using a sample of 600 velocity fields for the highest free-stream velocity.

\begin{figure}[!htb]
\begin{center}
\begin{tabular}{c}
\includegraphics[width=0.75\textwidth]{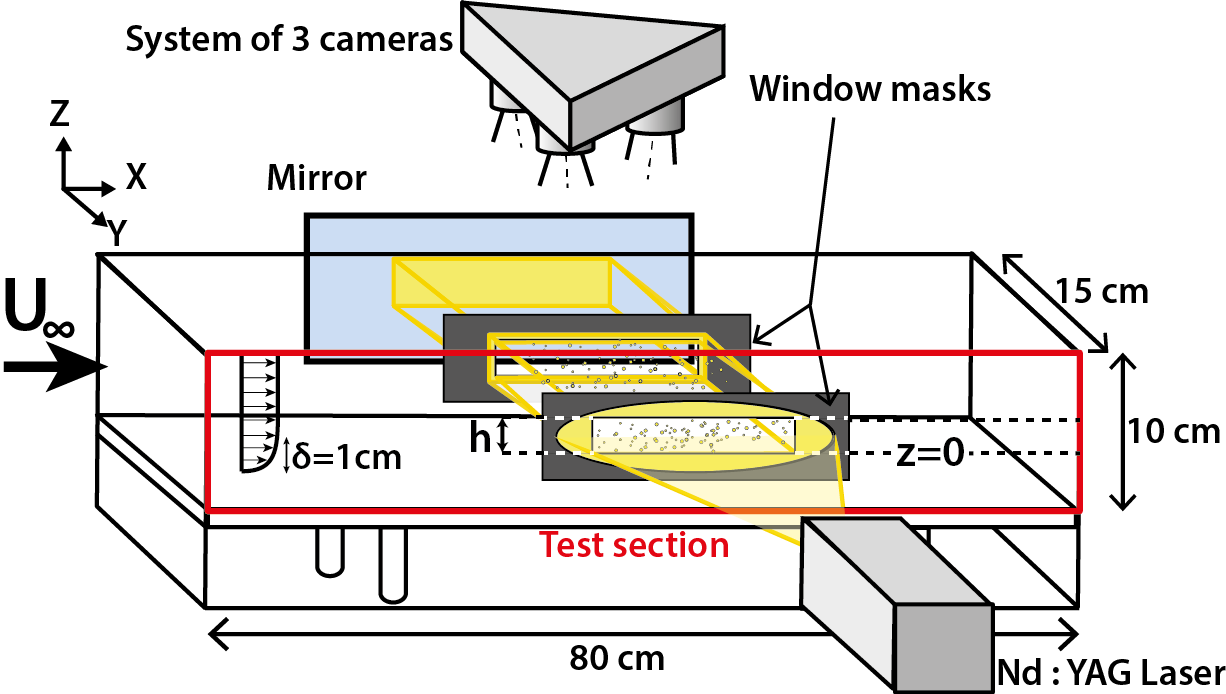}
\end{tabular}
\end{center}
\caption{Sketch of the test section. The measurement area is located in the test section of the channel (red frame in Fig. \ref{Fig1}) in the free-stream region.}
\label{Fig2}
\end{figure}

 The transparent test section is 80 cm long with a $10\times15$ cm$^{2}$ rectangular cross-section  (Fig. \ref{Fig2}).

The measurement volume used to investigate the screening effect is located in the middle of the channel (Fig. \ref{Fig2}). It is at least 2 cm away from each boundary layer $\delta_{BL}$ which are approximately 1 cm-thick. In this region, the velocity field is uniform with $U_{\infty} =$ 20 cm.s$^{-1}$. The decrease of processing efficiency are then independent of velocity gradient and can only be attributed to optical effects (optical screening, leveling of the intensity background). 

\begin{figure}[!htb]
\begin{center}
\includegraphics[width=0.5\textwidth]{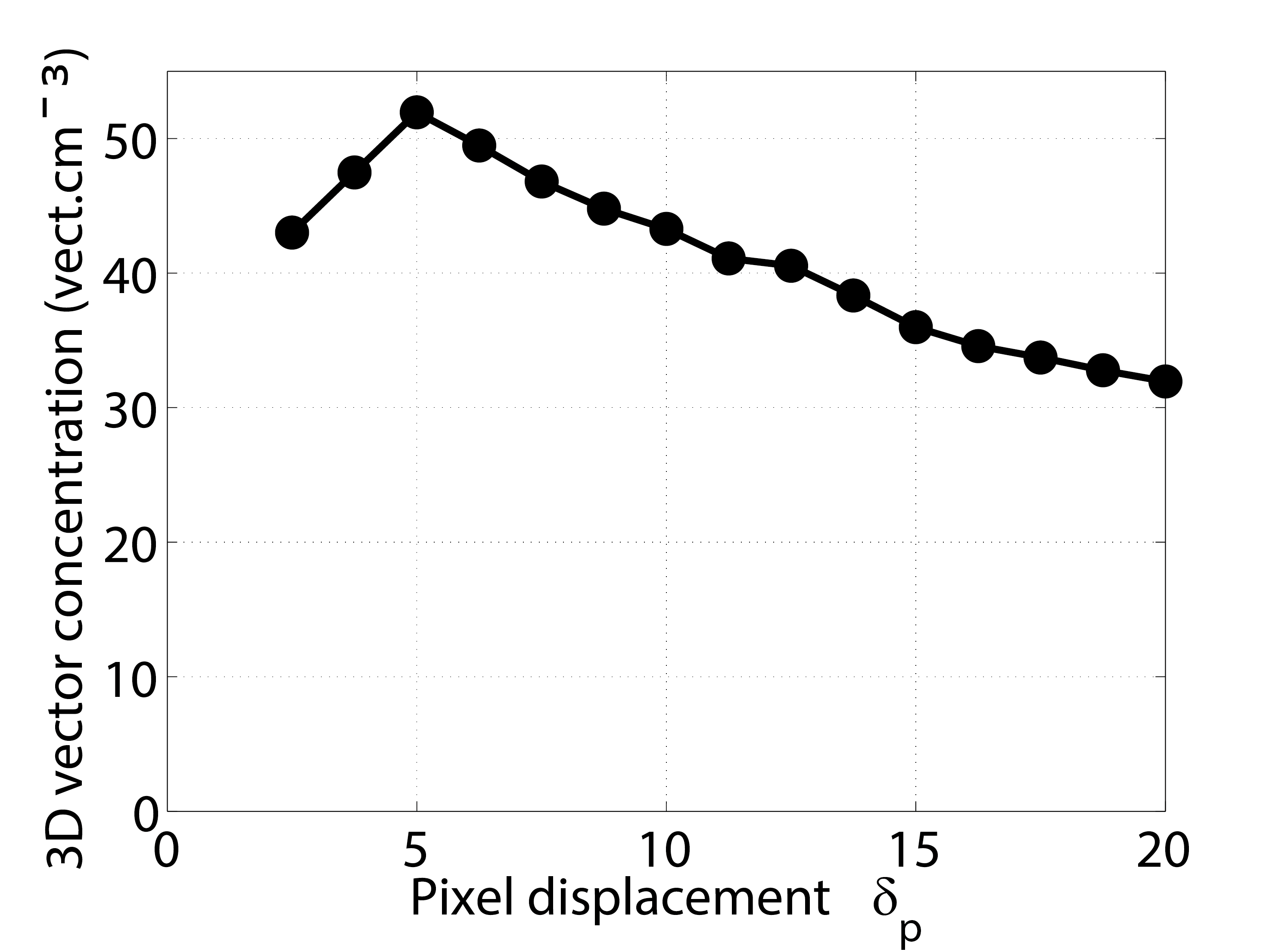}
\end{center}
\caption{Concentration of raw velocity vectors (vector.cm$^{-3}$) retrieved after the tracking step, as a function of the mean pixel displacement $\delta_p$ of the particles during $\delta t$.}\label{Fig3}
\end{figure}

The concentration of raw velocity vectors retrieved after the tracking step has been measured as a function of $\delta t$, as shown on Fig.~\ref{Fig3} for a 2 cm thick illumination volume and a moderate imaged particle concentration ($\approx\ 10^{-2}\ $ppp). An optimum is found for $\delta_p =$ 5 pixels. The same result was found  for other illumination thicknesses and other flow velocities, as long as the imaged particle concentration is kept moderate (below $4.5\times 10^{-2}\ $ppp, the screening phenomenon threshold). In the following, the time step between two snapshots is $\delta t = 3000\ \mu s$ which corresponds to a 5 to 6 pixels displacement of the particles.

The V3V probe is fixed above the water channel. Since its focal length $\mathrm{L_{f}=750}$ can not be changed (see subsection~\ref{Subsection:DDPIV} and Fig.~\ref{Fig6}), the whole hardware system can be precisely optimized to obtain the best accuracy (see also section \ref{uncertaintysection}). Therefore, following \citet{Pereira2002} recommendations, the measurement volume has been located  between $\mathrm{0.92 L_{f}}$ and $\mathrm{0.87  L_{f}}$.

Volumetric illumination is generated using a 200 $mJ$ pulsed Nd:YaG laser (Quantel)  and two perpendicular semi-cylindrical lenses.  Homogeneity of the illumination is critical for the particles detection. Using window masks, only the central part of the Gaussian illumination is used in order to obtain a quasi-uniform intensity in the measurement volume. 

A mirror located on the other side of the channel (Fig.~\ref{Fig2} and \ref{Fig4}a) reflects the laser beam into the test section to minimize the intensity attenuation along the laser beam axis (Beer-Lambert's law). Figure~\ref{Fig4}b shows  the spatial distribution of the detected and reconstructed 3D particles along the spanwise direction $y$ with and without the mirror. Without mirror, more particles are found on the laser side. When a mirror is used the number of successfully detected particles along the channel becomes approximately constant along the channel width. The mirror clearly allows for a better detection of the particles along the entire width of the channel and proves to be a simple and effective way to homogenize the intensity distribution along the beam axis.

To avoid screening effects outside the region of interest of the measurement volume, window masks have been used to illuminate only the region where the velocity field has been measured (Fig.~\ref{Fig2} and \ref{Fig4}a).  Window masks with various heights $h$ (1 cm $< h <$ 4 cm) have been used to study the influence of the thickness of the illumination volume.  The illumination properties (laser intensity, position width and height of the laser volume) are unchanged. The location of the lower part of the measurement volume is the same for all the measurements so that the upper part of the measurement volume gets closer to the cameras for increasing $h$ (Fig.~\ref{Fig8}).

Fifty images were taken for each configuration to ensure proper convergence of the mean values of the processing parameters (defined in the following section) like the different concentrations of particles or raw velocity vectors.

\subsection{Definition of the measurement volume}

\begin{figure}[!htb]
\begin{center}
\begin{tabular}{cc}
      \includegraphics[width=0.55\textwidth]{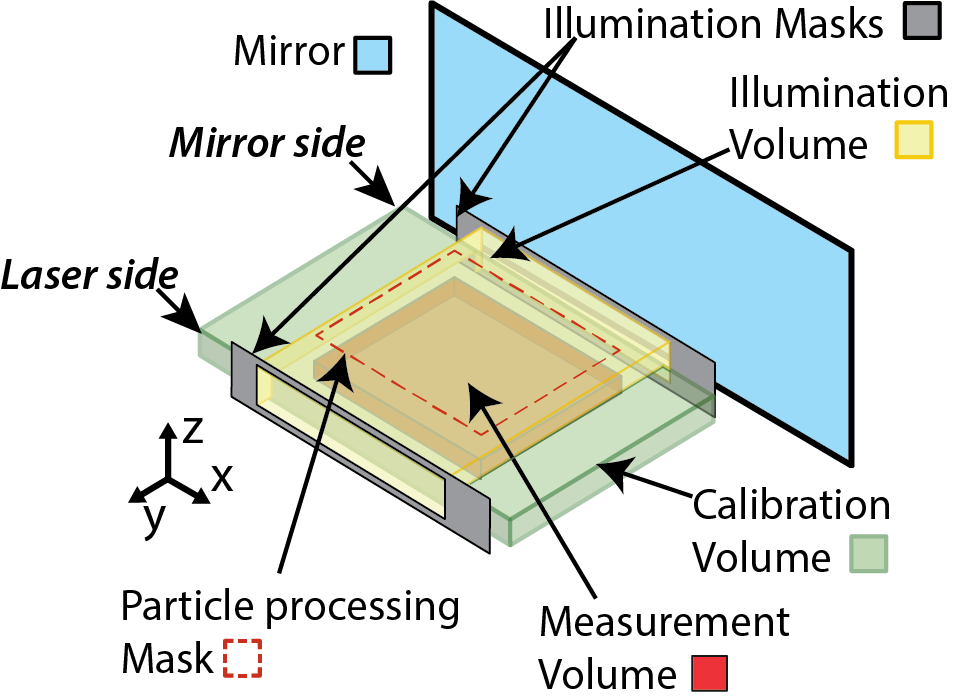}&   \includegraphics[width=0.45\textwidth]{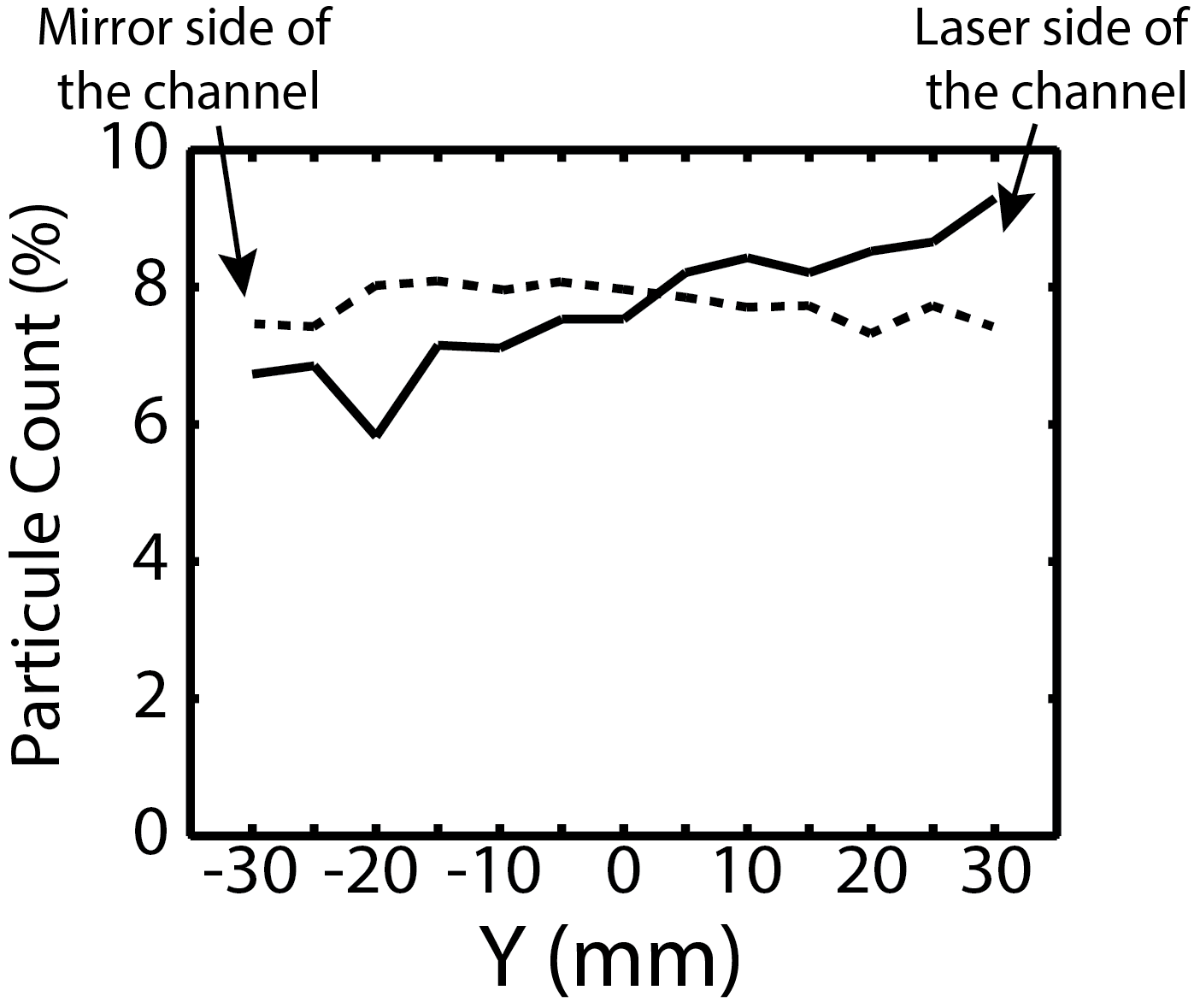}\\
a)&b)
\end{tabular}
\end{center}
  \caption{a) The measurement volume is the intersection between the illumination volume, the calibrated volume and the particle processing mask volume. b)  Spatial repartition of 3D detected particles along the channel width  with a mirror ($--$) and without a mirror ($-$).}\label{Fig4}
\end{figure}

To properly define the measurement volume, three distinct volumes have to be considered:
\begin{itemize}
 \item the calibrated volume, where the three-dimensional position of a particle can be computed. 
\item the illumination volume, defined by the window masks and the width of the channel.  
\item  the volume defined by the particle processing masks field of view. For clarity, only one particle processing mask is displayed on Fig. \ref{Fig4}a. In practice one particle processing mask is defined for each camera (see for instance the camera images superposition on Fig. \ref{Fig6}).
\end{itemize}

The measurement volume corresponds to the intersection between the illumination volume, the calibrated volume and the particle processing mask volume.  
Special care has been taken to adjust both the illumination and calibrated volumes to maximize the size of the measurement volume and to optimize the efficiency of the processing. Indeed, if these two volumes are different, either a part of the calibrated volume is not illuminated, or particles outside the calibrated volume are illuminated. Without illumination, particles can not be detected. Without calibration, particles can not be reconstructed. In this case, they are not only useless, they also lower  the image quality. Located on the front of the measurement volume, these undesired particles hide useful particles and make their detection more difficult.

\subsection{Seeding}

The choice of the particle size is very important. In the present study, Polyamide 50$\mu$m Dantec Dynamics particles (size distribution between 30 and 70 $\mu$m) were used.  Their density is $\rho_p =$1.03 g.cm$^{-3}$. The sedimentation velocity is $4.10^{-3}$ cm.s$^{-1}$ and can therefore be neglected. 
\begin{figure}[!htb]
\begin{center}
\includegraphics[width=0.2\textwidth]{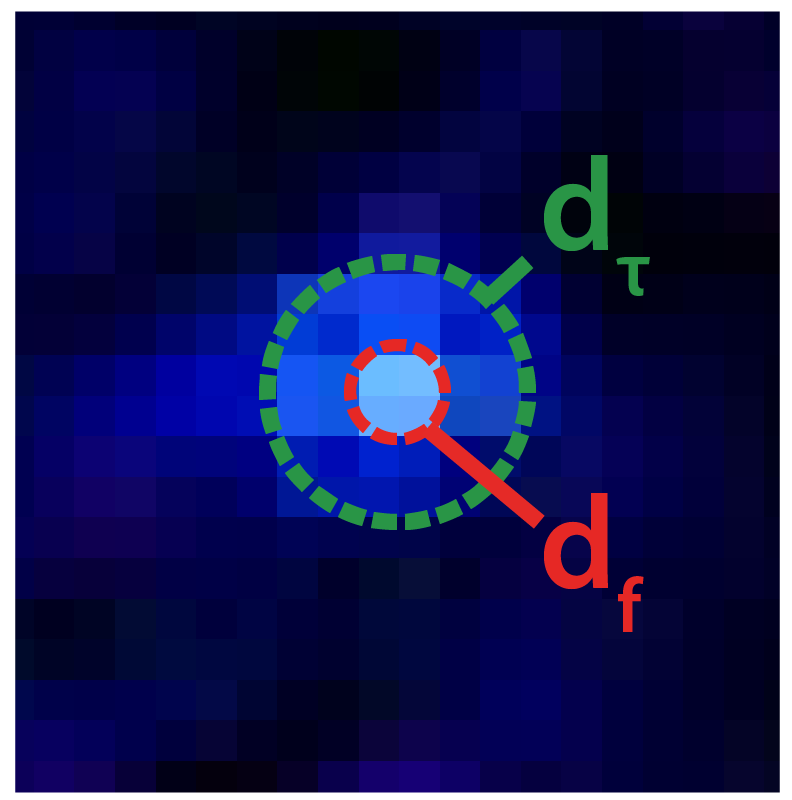}
\end{center}
\caption{Single particle image. The effective particle image diameter $d_\tau$, and the gaussian-fitted diameter $d_f$ are represented. The background noise intensity is around $\approx6-9$.}\label{Fig5}
\end{figure}

Knowing the particle diameter $d_p$ and the optical characteristics of the setup, it is possible to compute the diameter of the effective particle image $d_\tau$ seen by the camera sensor (\citet{Olsen2000}). Taking into account the magnification factor M $\approx$ 0.1, the $d_p=30-70$ $\mu m$ size distribution leads to a $d_\tau\approx3-7$ pixels distribution of the effective particle images (Fig. \ref{Fig5}). This is a good size distribution to ensure a good estimation of the intensity gradients and to determine precisely the particle center. This is a crucial step which influences the whole technique uncertainty. Indeed, an error on the 2D particle center directly leads to an error on the 3D particle reconstruction and then on the evaluation of the particle displacement.

This effective diameter distribution of the particle images leads to a diameter distribution of gaussian fit profiles $d_f$ (Fig. \ref{Fig5}). This distribution ranges from  $d_f=0.8$ to 2 pixels, with a peak around 1 pixel (cf the $\mathrm{C_{m1}}$ curve in the Fig. \ref{Fig13}). In the end, this close-to-one value of the fitted diameter $d_f$ ($\Leftrightarrow  d_\tau\approx 3$ pixels) is the lower bound below which the particle size is too small and would induce errors in the estimation of the particle center. Therefore, this particle size allows for the smallest useful particle image diameter, and then the largest possible imaged concentration of detected particles per image. 

It should be noted that, since the V3V probe always has to be placed at the same distance from the measurement volume to optimize the measurement uncertainty (fixed focal length magnification), the magnification factor is always the same (M$\approx$0.1). Taking into account the pixel size, and the detection algorithm of the system, the 30$-$70 $\mathrm{\mu}$m particle diameters are not only a relevant choice of particle size for this study. Due to the software and hardware constraints, it also corresponds to an optimal choice for V3V measurements.

\section{Volumetric velocimetry}
 The V3V probe is made up of three 4 MP (2048$\times$2048 pixels) 12 bits double-frame cameras. The pixel size is 7.4 $\times7.4 \mu$m$^2$.  The maximum acquisition frequency is  $f_{ac}=$ 7.25 Hz.  This system is based on the principle of Defocusing Digital Particle Image Velocimetry (DDPIV) to retrieve the 3D particle positions and uses a 3D-PTV algorithm to track 3D particles in space (Relaxation method).
 
\subsection{DDPIV principle}\label{Subsection:DDPIV}
\begin{figure}[!htb]
\begin{center}
\includegraphics[width=0.9\textwidth]{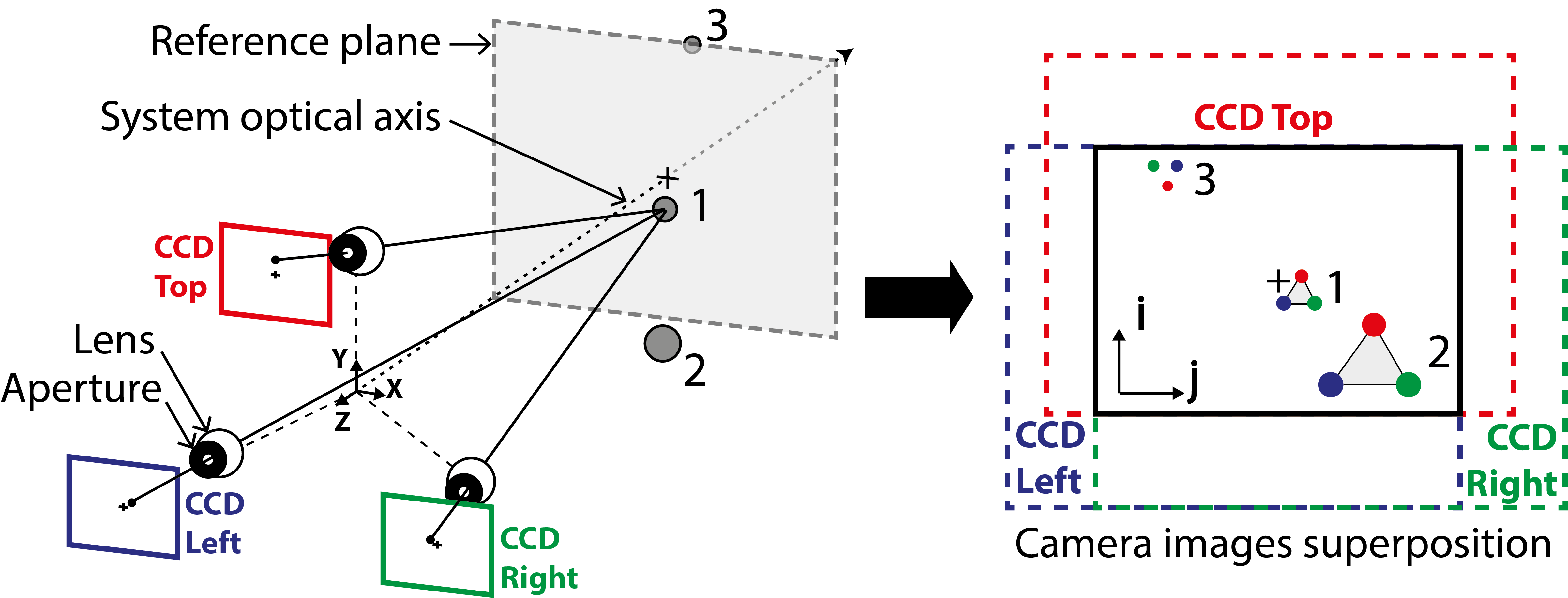} \\
 \caption{DDPIV principle applied on a 3D three cameras triangular configuration. Definition of triplets used to evaluate the location of the particles in the volume.}\label{Fig6}
\end{center}
 \end{figure}

This method was for the first time described by \citet{willert1992} and was applied later by \citet{Pereira2000} for a configuration with three cameras. Fig. \ref{Fig6} shows how a set of three cameras visualize each particle as a triangular pattern of dots, also named triplets. The triplet size is used to evaluate the position of the particles along the optical axis. In the focal plane the triplets collapse as single dots. The performances and limitations of the DDPIV method have already  been thoroughly studied (\citet{Graff2008,Pereira2006,Pereira2002,Kajitani2005,Grothe2008}).

To retrieve the axial position of a particle using the triplet size, a calibration of the measurement volume is necessary. Pictures of a calibration plate (with 5 mm regular grid) are taken for different positions along the optical axis. During the calibration, the plate is moved along the optical axis of the system using a $\Delta y=1$ mm increment step. The  precision of the calibration determines the error on the estimation of the particle centers during the 3D particle reconstruction step. The precision depends on mainly four parameters: the strict orthogonality of the calibration plane with the system optical axis,  the error made on the displacement increments,  the precision of the calibration grid and the homogeneity of the illumination.  A typical value of this reconstruction error for a good calibration is approximately of 0.2 pixel for each camera (provided by TSI). This value has been obtained with our calibration for the present experiments.

\subsection{Processing}\label{sec:Processing}
First, a pre-processing is carried out to improve the contrast between the background and the particles. A background image is generated for each camera from the time-series. For each pixel, the minimum intensity value on the whole time-series is kept. Therefore, the background image contains every motionless elements (flat plate, walls reflexions, bubbles, ...) that could lower the contrast between the particles and the background. It is then removed from every picture of the time series. This preprocessing step is important to improve particles detection. 

\begin{figure}[!htb]
\begin{center}
\includegraphics[width=\textwidth]{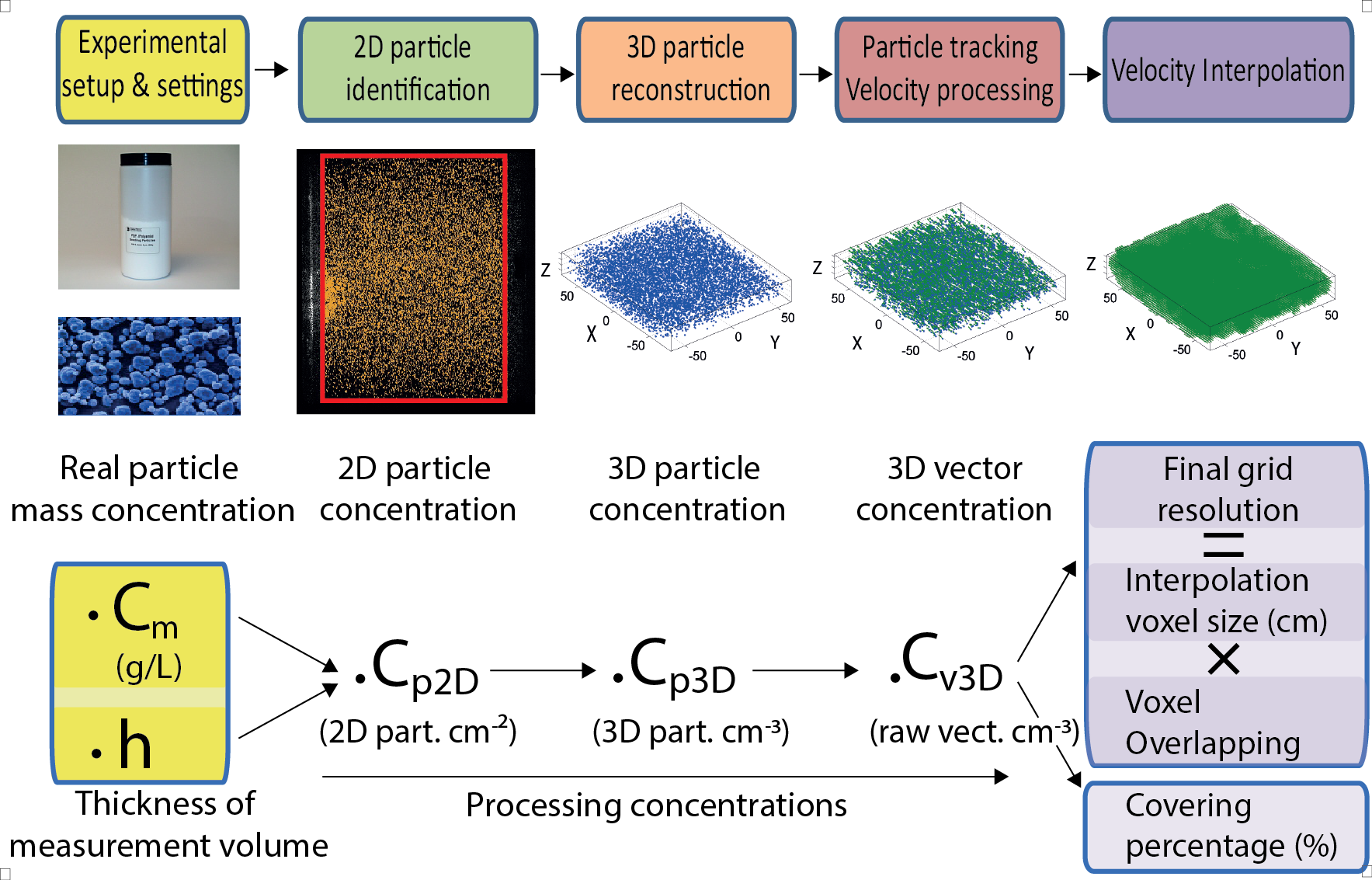}
\end{center}
\caption{Starting from the real particle mass concentration, the different concentrations used at each processing step are defined, together with a schematic diagram of the principal processing steps of the TSI software.}\label{Fig7}
\end{figure}

Then, four processing steps are necessary to compute a 3D interpolated velocity field out of the set of 6 raw pictures. The main steps are explained in the following and are summarized on Fig.\ref{Fig7}.

\textbf{The 2D particle identification step} 
In this first step the intensity peaks are detected in the regions of interest inside the particle processing masks (the red windows in Fig. \ref{Fig4}a and \ref{Fig7}) of the three pairs of images (one for each camera at time $t$ and one for each camera at time $t+\delta t$). 
This peak detection uses an intensity threshold  (50 on a 4096 gray scale) leads to a signal to noise ratio $SR\approx5-8$).
Due to the pre-processing, the maximum background noise ranges between 6 to 9 (signal to noise ratio $SR\approx5-8$). 
The gaussian-fitted diameter $d_f$ ranges  between 0.5 and 3 pixels. A  70\% particle overlap  is chosen. Using these parameters each particle can be fitted by a Gaussian intensity profile through an iterative scheme on neighboring pixels (Levenberg-Marquardt algorithm). Each 2D particle is then defined by its center and its fitted diameter $d_f$. A detailed description of these algorithms can be found in \citet{Pereira2002}. \\

\textbf{The 3D particle reconstruction step}  
In this second step, particles identified on one camera are matched with the corresponding particles seen by the other two cameras. Fine and coarse search tolerances (respectively  0.5 and 1 pixel) and the system calibration are used in this matching process to retrieve  a triplet of coordinates for each particle (Fig. \ref{Fig6}).  The 3D particle positions are then reconstructed using the DDPIV relations on the triplets. More details can be found in \citet{Pereira2002,Troolin2010}.   Since the V3V system is a multi-view vision system,  ''ghost particles'' \citep{Graff2008} are retrieved after this reconstruction step. They can be either real particles if two \lq\lq{}real\rq\rq{} particles are perfectly overlapped in one camera image or ''false particles'' otherwise. The number of false particles obviously increases with increasing 2D concentrations of particles in the image. To avoid the screening effect and still use high 2D particle concentrations, only  the best triplets with respect to the reconstruction validity criteria are kept. All ghost particles are dismissed. 

The raw yield $R_{ry}$ measures the efficiency of the 3D particle reconstruction step. It corresponds to  the ratio of the number of valid particles in 3D (3D particles count: $PC_{3D}$) over the number of detected particles in 2D (2D particles count: $PC_{2D}$): $R_{ry} = PC_{3D}/PC_{2D}$.  It measures the efficiency of the reconstruction step and ranges between 0 (no rebuilding) and 1 (no loss).\\

\textbf{The tracking step (or velocity processing step)} 
In this step, a particle located somewhere in the volume at time $t$  is tracked at time $t+\delta t$, leading to a first order estimate of its instantaneous velocity $\delta \overrightarrow{x}/\delta t$. The tracking algorithm used in this study is the relaxation method (\citet{Pereira2002,Baek1996,Pereira2006}). Finally, a 3D raw velocity vectors field, randomly distributed in the measurement volume, is obtained.

\textbf{The velocity interpolation step} 
During this last step, the random distribution of velocity vectors in the volume is interpolated on a volumetric interpolation grid in order to obtain a spatially homogeneous velocity field.
The final grid is defined using an interpolation grid size (voxel size), an overlapping percentage of the voxels (voxel overlapping) and a smoothing factor which takes into account the influence of the neighboring voxels (not used in our measurements). For instance, a 4 mm voxel size combined with a 75$\%$ voxel overlapping produce a spatial resolution of 1 mm for the final velocity field. 
 This last step is achieved using a Gaussian-weighted interpolation algorithm whose Gaussian size matches the grid size interpolation (in the case without smoothing).  

Depending on the spatial homogeneity of the initial distribution of particles and depending on the concentration of raw vectors obtained after the step of velocity processing, the interpolation proves to be possible or not in each voxel.  If there are not enough raw velocity vectors in the vicinity of a node, the interpolation is not considered as successful. \citet{Pereira2002} show that a seven neighboring raw velocity vectors per voxel is a minimum requirement to obtain a good interpolation and to keep the error low over a range of displacements. 

As a consequence, there can be holes in the interpolated fields. The percentage of the nodes of the final 3D grid successfully interpolated is called \emph{covering percentage} and is noted $R_{\% cover}$. It is the ratio between the number of nodes where the velocity is successfully interpolated over the total number of nodes of the 3D grid.  For example, if on a 10$\times$10 grid (for a 2D case) the interpolation is successfully performed on 91 nodes, it leads to $R_{\% cover}=91\%$.
The objective is then to reach a covering percentage as close a possible from 100 $\%$ to obtain good instantaneous 3D velocity fields.  

 \subsection{Uncertainty of volumetric velocimetry measurements}\label{uncertaintysection}
This question has been well-studied in previous works  \cite{Pereira2006,Pereira2002,Grothe2008,Graff2008}.
The uncertainty in the volumetric PTV measurements is due to the difficulty in defining properly the coordinates of the particles  and therefore determining the centroid of corresponding triplets. The error on particle coordinates in our experiments is lower than 0.2 pixel. Combining pixel displacement errors with processing errors, \citet{Pereira2002} determine an uncertainty level in the (X, Y) plane  (cf Fig. \ref{Fig6}) for the velocity measurements of $\pm$1\%,  whereas the error on the depth displacement is four to six times higher. 

\section{Influence of the concentration of particles, thickness of the measurement volume and screening}	
 \subsection{Definition of the optical screening}
As mentioned previously, a particle can hide another one, preventing its detection during the step of 2D particle identification and making impossible the reconstruction of the 3D particle position, even if this particle is correctly detected on two of the three cameras. In the following this phenomenon is called \emph{optical screening if the statistical properties of the  distribution of detected and reconstructed particles are significantly affected}. Fig. \ref{Fig9} illustrates how this phenomenon sets up, while Fig. \ref{Fig10} shows its effect on the distribution of the detected and reconstructed particles along the optical axis. One can already see that the optical screening happens for high concentrations of the particles in the images. Intuitively, one can guess that a compromise has to be found between the particles concentration in the flow and the thickness of the illumination volume. Indeed, both parameters influence the visual particle concentration in the camera images (Fig. \ref{Fig8}). The objective of this section is to quantify the effect of each parameter on the optical screening.

\subsection{Experimental parameters}
The optical screening becomes more important for increasing particle concentrations and increasing illumination thicknesses. Therefore these two parameters have been varied.

In order to study the influence of the particle concentration on screening, water was first filtered using  a $5 \mu$m filter. 
The mass concentration in particles $\mathrm{C_{m}}$ was then gradually increased from $\mathrm{C_{m}\approx0}$ to $\mathrm{C_{m}\approx5.6.10^{-3}\ }$g.L$^{-1}$, with $\mathrm{\Delta C_{m} \approx 1.5.10^{-3}\ }$g.L$^{-1}$ increments.

\begin{figure}[htb!]
\begin{center}
\begin{tabular}{cc}
\includegraphics[width=0.45\textwidth]{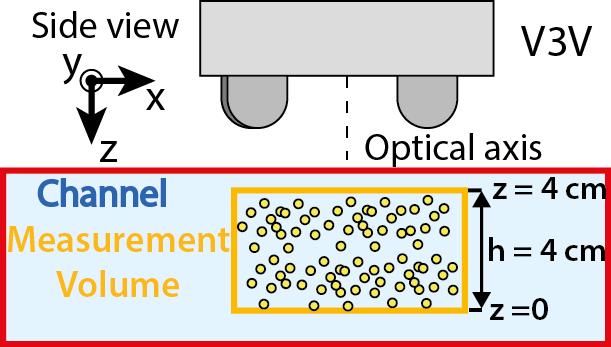} &
\includegraphics[width=0.45\textwidth]{Fig8b}\\
a)&b)
\end{tabular}
\caption{Variation of the thickness $h$ of the measurement volume: a) $h=$2 cm; b) $h=$4 cm.}
\label{Fig8}
\end{center}\end{figure}

Five measurement volume thicknesses were used: $h=$ 1, 1.5, 2, 3 and 4 cm. The positions of the  $h =$ 2 cm and $h =$ 4 cm  measurement volumes are illustrated on the Fig.\ref{Fig8}.

\subsection{Illustration of the screening phenomenon}
Four different concentrations, corresponding to four characteristic states, are defined for a 4-cm-thick measurement volume:
\begin{itemize}
\item  $\mathrm{C_{m1}\approx 1.5\ 10^{-3}\ }$g.L$^{-1}$ : Minimum seeding. \\Imaged concentration $Cp_{2D}\approx 10\ part.cm^{-2}=6\cdot 10^{-4}\ ppp$.
\item  $\mathrm{C_{m10}\approx 1\ 10^{-2}\ }$g.L$^{-1}$ : Optimal seeding. \\$Cp_{2D}\approx 730\ part.cm^{-2}=4.5\cdot10^{-2}\ ppp$.
\item  $\mathrm{C_{m20}\approx 2.3\ 10^{-2}\ }$g.L$^{-1}$: Critical seeding: settling of an optical screening.\\$Cp_{2D}\approx 1000 \ part.cm^{-2}=5.95\cdot10^{-2}\ ppp$.
\item  $\mathrm{C_{m40}\approx 5.6\ 10^{-2}\ }$g.L$^{-1}$: Final over-saturated seeding. \\$Cp_{2D}\approx 1200 \ part.cm^{-2}=7.1\cdot10^{-2}\ ppp$.
\end{itemize}

\begin{figure}[htb!]\begin{center}
\begin{tabular}{|c|c|c|c|}
\hline
a)&b)&c)&d)\\
\includegraphics[width=0.22\textwidth]{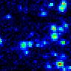} & 
\includegraphics[width=0.22\textwidth]{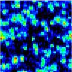} & 
\includegraphics[width=0.22\textwidth]{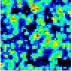} & 
\includegraphics[width=0.22\textwidth]{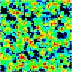}\\
e)&f)&g)&h)\\
\includegraphics[width=0.22\textwidth]{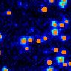} & 
\includegraphics[width=0.22\textwidth]{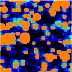} & 
\includegraphics[width=0.22\textwidth]{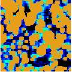} & 
\includegraphics[width=0.22\textwidth]{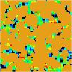}\\\hline
$C_{m1}$&$C_{m10}$&$C_{m20}$&$C_{m40}$\\\hline
\end{tabular}\end{center}
\caption{Illumination thickness $h=$ 4 cm. Zoom over 75$\times$75 pixels square images. \newline  a, b, c, d) Instantaneous images for $\mathrm{C_{m1}}$, $\mathrm{C_{m10}}$, $\mathrm{C_{m20}}$ and $\mathrm{C_{m40}}$. e, f, g, h) Same images where the detected particles and their computed diameter is shown using filled orange circles.}\label{Fig9}
\end{figure}

The Fig. \ref{Fig9}a, \ref{Fig9}b, \ref{Fig9}c, \ref{Fig9}d show typical images obtained for the four concentrations previously defined ($\mathrm{C_{m1}, C_{m10}}$, $\mathrm{C_{m20}}$ and $\mathrm{C_{m40}}$). The Fig. \ref{Fig9}e, \ref{Fig9}f, \ref{Fig9}g, \ref{Fig9}h show the same images with the position and the diameter of the detected particles shown as orange discs.

These images show the increasing visual density of particle images for increasing mass concentration. Fig. \ref{Fig9}a shows a very low mass particle concentration which results in a low imaged concentration, even if the illumination volume is 4 cm-thick. Clearly, no optical screening takes place in this case. On the opposite, one can see in the Fig. \ref{Fig9}d that the image is clearly saturated by particle images. Neighboring particles images overlap each other and totally mask the background. This moving bed of particles totally hides the particles in the back of the measurement volume. It is a good illustration of the screening phenomenon.

\begin{figure}[htb!]

\begin{center} \begin{tabular}{cc}
\includegraphics[width=0.5\textwidth]{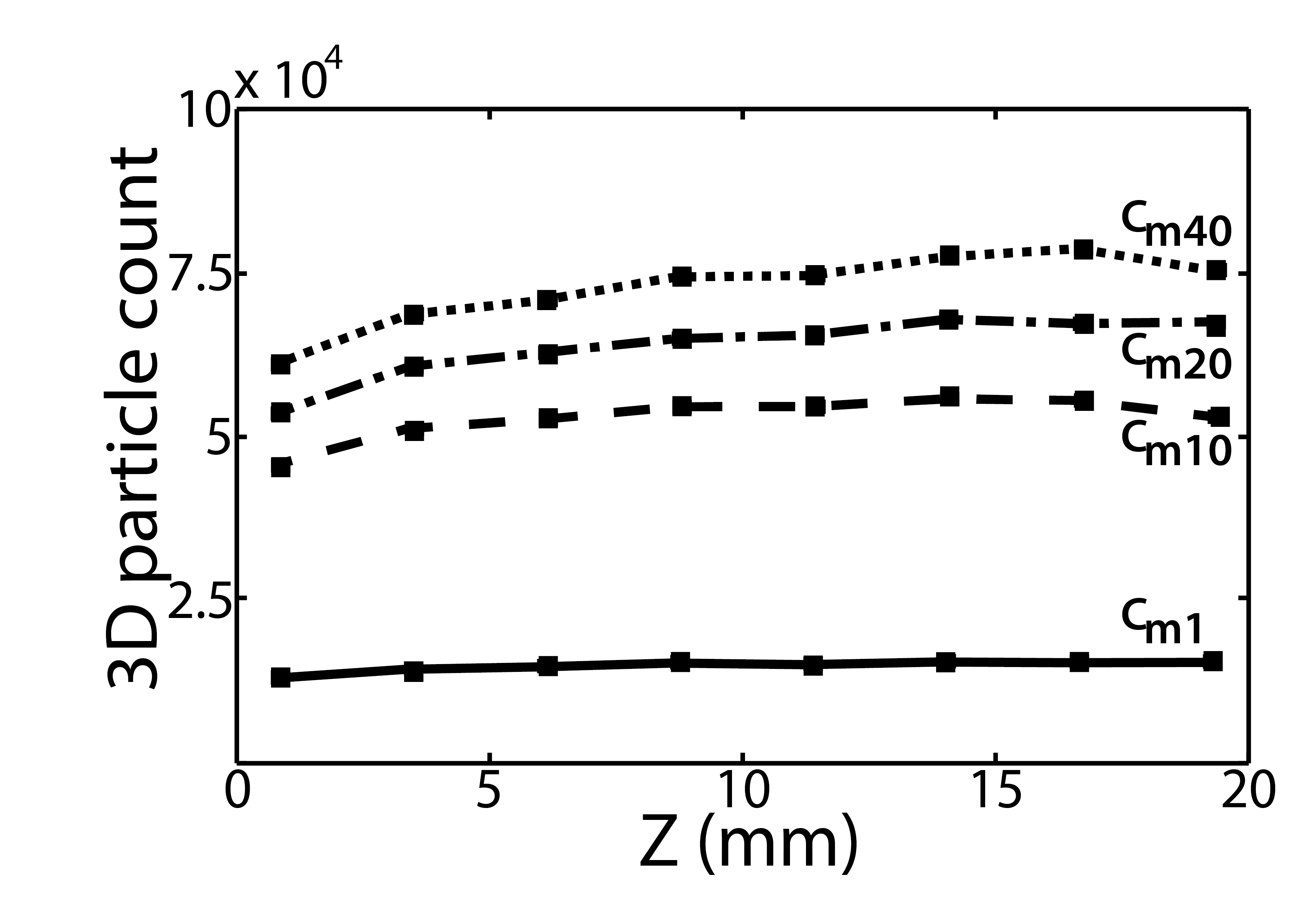}&
 \includegraphics[width=0.5\textwidth]{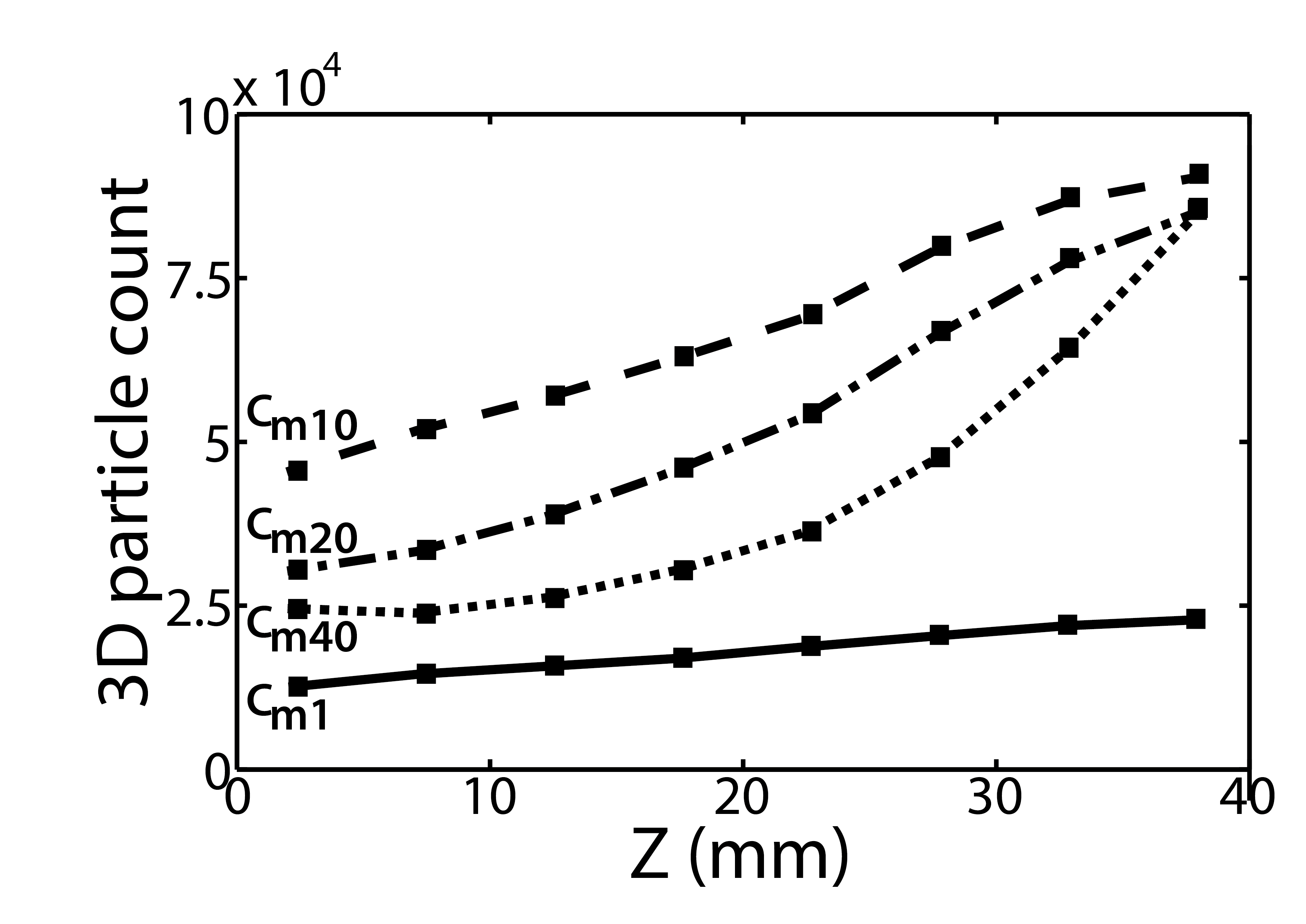} \\
a)&b)\\
\includegraphics[width=0.5\textwidth]{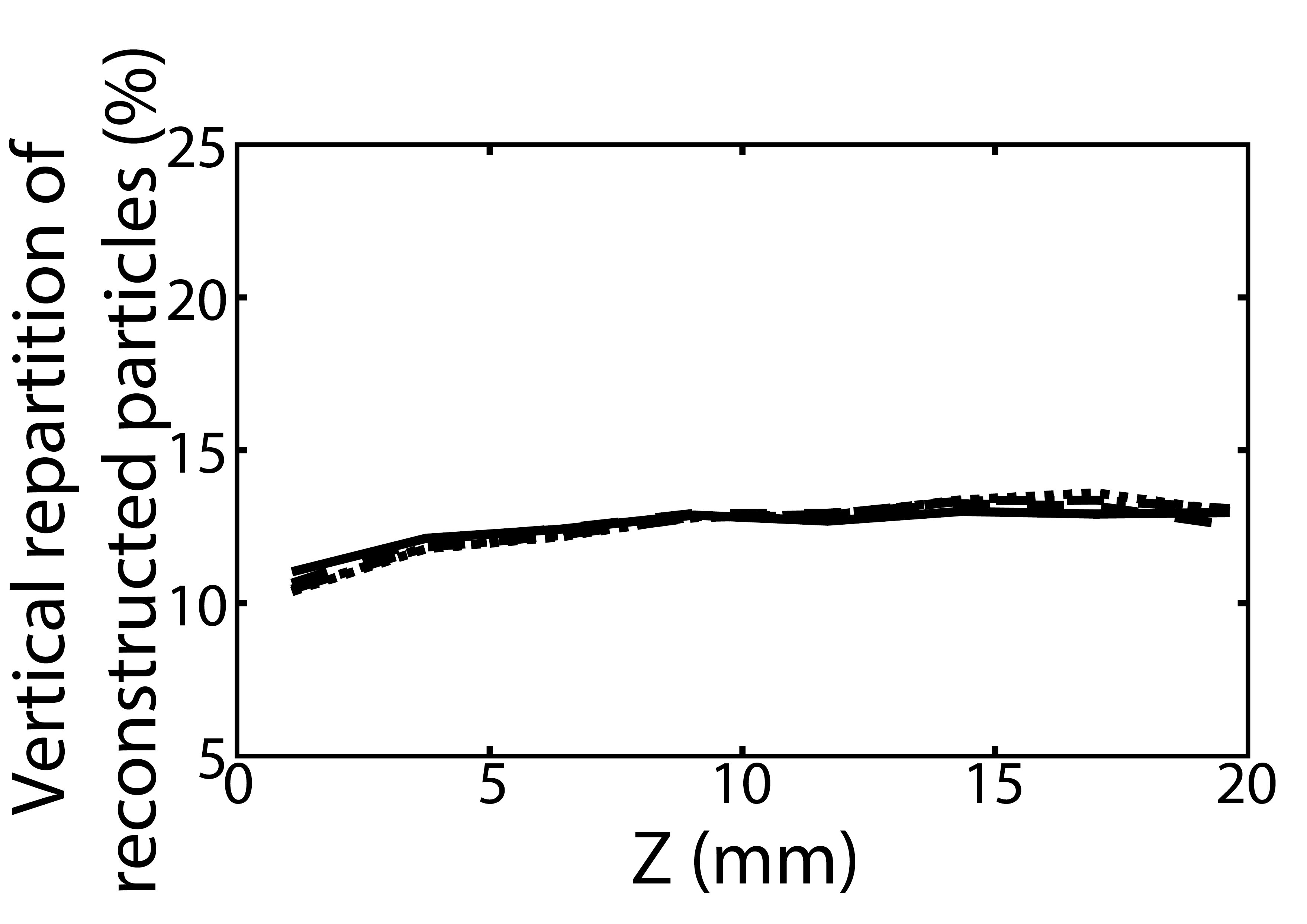}& 
\includegraphics[width=0.5\textwidth]{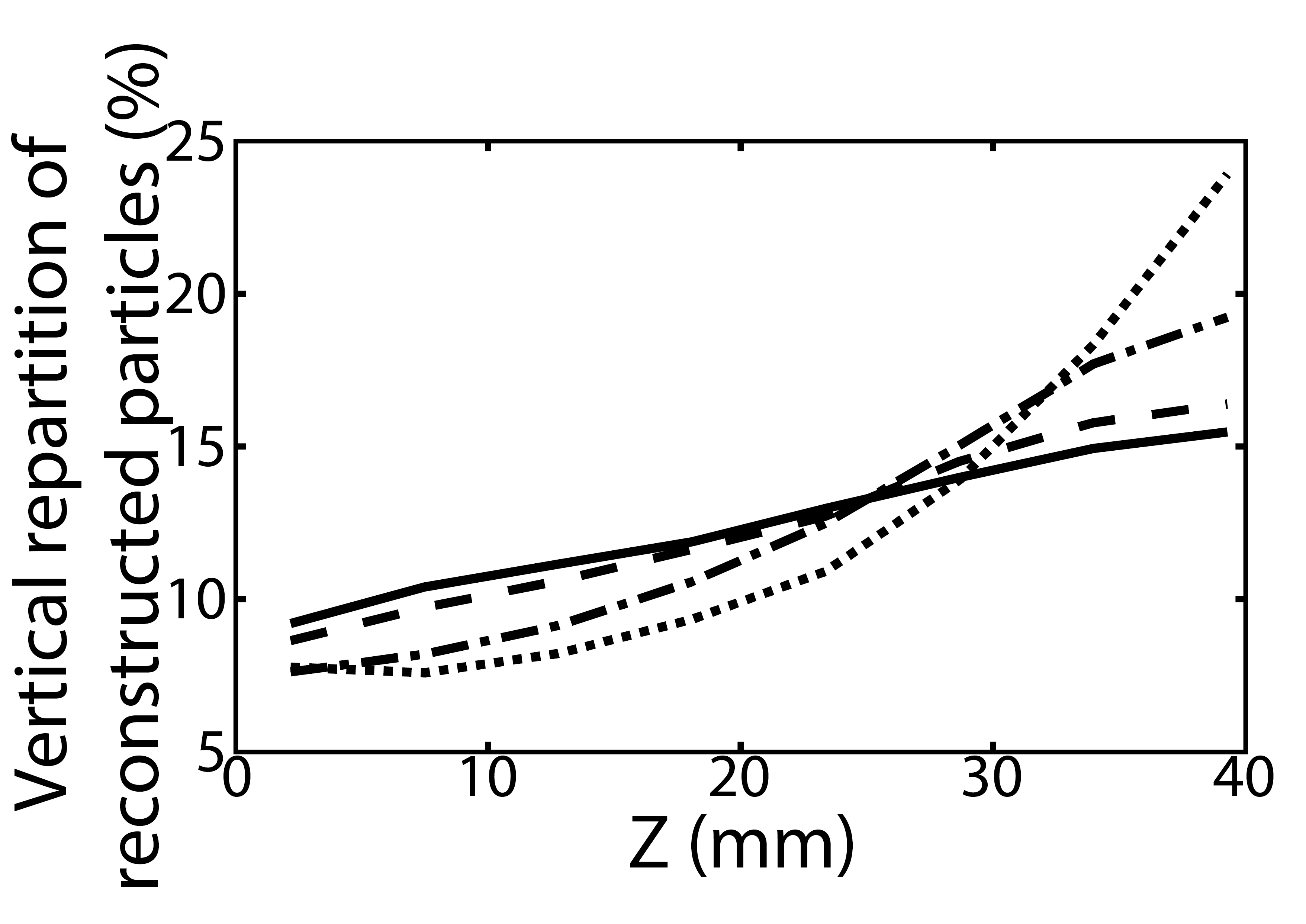} \\
c)&d)
\end{tabular}
\caption{a) Vertical repartition (in particle count) of reconstructed 3D particle positions for an illumination thickness $h=$ 2 cm. b) Vertical distribution (in particle count) of reconstructed 3D particle positions for an illumination thickness h$=$ 4 cm. c) Vertical repartition (in percentage) of reconstructed 3D particle positions for an illumination thickness h$=$ 2 cm. d) Vertical distribution reconstructed 3D particles positions for an illumination thickness $h=$ 4 cm.\newline
Mass concentration: $\mathrm{C_{m1}}$ ($-$), $\mathrm{C_{m10}}$ ($---$), $\mathrm{C_{m20}}$ ($- \cdot - $), $\mathrm{C_{m40}}$ ($\cdots$).}
\label{Fig10}
\end{center}\end{figure}

The Fig. \ref{Fig10}(a, b, c, d) demonstrate the role of the thickness of the measurement volume in triggering the optical screening. The distribution along the optical axis ($Z$ axis) of the 3D particles reconstructed is shown for two illumination thicknesses: $h = 2$ cm (Fig. \ref{Fig10}a and \ref{Fig10}c) and $h = 4$ cm (Fig. \ref{Fig10}b and \ref{Fig10}d), and for the same concentrations. The Fig. \ref{Fig10}a and \ref{Fig10}b show the amount of 3D particles detected in a fluid layer centered around a given $Z$ position. The Fig. \ref{Fig10}c and \ref{Fig10}d show the same repartition normalized by the total number of particles detected in the whole measurement volume for each concentration. For an homogeneous distribution of particles and homogeneous illumination over the measurement volume,  a flat distribution profile is expected. This is the case for the 2-cm-thick illumination volume. The total number of reconstructed particle increases with the mass concentration  (Fig. \ref{Fig10}a), but the repartition along the optical axis remains the same  (Fig. \ref{Fig10}c). A slight slope is still noticed but it does not correspond to any screening effect since it does not depend on the mass concentration.\\ 

On the opposite, the vertical distribution of reconstructed particles along the $Z$ axis depends strongly on the mass concentration for the 4-cm-thick illumination volume (Fig. \ref{Fig10}b and \ref{Fig10}d). For small concentrations of particles ($\mathrm{C_{m1}}$ and $\mathrm{C_{m10}}$), the distribution profiles in Fig. \ref{Fig10}c are the same as the ones in Fig. \ref{Fig10}d. As a consequence it is due to the properties of the volumetric system itself (calibration + detection + reconstruction) rather than the optical screening phenomenon. For $\mathrm{C_{m10}}$ the number of 3D particles reconstructed is maximum (Fig. \ref{Fig10}b). For $\mathrm{C_{m20}}$ and $\mathrm{C_{m40}}$, the distribution profiles become strongly nonlinear: particles are much more successfully detected and reconstructed in the front than in the back of the measurement volume (Fig. \ref{Fig10}d). For $\mathrm{C_{m20}}$ and $\mathrm{C_{m40}}$, the detection and reconstruction steps are less efficient and lead to less reconstructed particles than the  $\mathrm{C_{m10}}$ case (Fig. \ref{Fig10}b).
All these results can be explained by the optical screening phenomenon which strongly influences the performances of the detection and reconstruction algorithms due to high concentration.

\subsection{Definitions of processing concentrations}
In order to study the effect of particle mass concentration on each processing step, several new concentrations (Fig. \ref{Fig7}) have to be introduced:

\begin{itemize}
 \item   The concentration of 2D particles per $cm^2$ detected in each image, noted $\mathrm{Cp_{2D}}$ :
$$Cp_{2D}= \frac{PC_{2D}}{L_{x}\cdot L_{y}}= \frac{PC_{2D}}{Pix_{count} \cdot C_{f}^{2}},$$ 
where $L_{x}$ and $L_{y}$ are the dimensions of the particle processing mask (Fig.  \ref{Fig4}), $Pix_{count}$ is the number of pixels included inside the particle processing mask, $PC_{2D}$ is the number of 2D particles  detected in the mask and $C_{f}$ is the conversion factor $\mu m/pixels$.  $C_{f}$ is the average of the local $Cf_{local}$ of the extreme planes defining the volume of illumination (Table \ref{tableLpxlocal}).  
This concentration is directly related to the screening phenomenon. 
 \item   The concentration of successfully reconstructed 3D particles per $cm^3$: $$Cp_{3D}\quad=\quad \frac{PC_{3D}}{L_{x} \cdot L_{y} \cdot L_{z}}$$ where $PC_{3D}$ corresponds to the number of particles detected and reconstructed  in the measurement volume. This quantity noted $\mathrm{Cp_{3D}}$ controls the efficiency of the tracking step  and the density of raw velocity vectors.
 \item   The concentration of raw velocity vectors per $cm^3$ obtained after the third processing step can be defined as: $$Cv_{3D}= \frac{VV_{count}}{L_{x} \cdot L_{y} \cdot L_{z}}$$ where $VV_{count}$ corresponds to the number of raw velocity vectors obtained in the measurement volume.
 \end{itemize} 

\begin{table}[!h]
\begin{center} 
\begin{tabular}{| l || c | c | c | c | c | c || }
 \hline 
\small{$Z_{plane}$ (cm)} & 0 & 1 & 2 & 3 & 4\\     
\hline 
$Cf_{local}$  ($\mu$m.pixel$^{-1}$)& 84.73  & 84.23    & 83.42  & 82.79  & 82.28  \\ 
  \hline
 \end{tabular}\caption{Local conversion factors $\mu$m to pixel at different positions of the plane along the optical axis $Z_{plane}$. It is obtained by calibration of the measurement volume. }\label{tableLpxlocal}
 \end{center}
\end{table}

\subsection{Influence of $\mathrm{C_{m}}$ on $\mathrm{Cp_{2D}}$, $\mathrm{Cp_{3D}}$ and $\mathrm{Cv_{3D}}$ }

\begin{figure}[htb!]
\begin{center} 
\begin{tabular}{cc}
\includegraphics[width=0.5\textwidth]{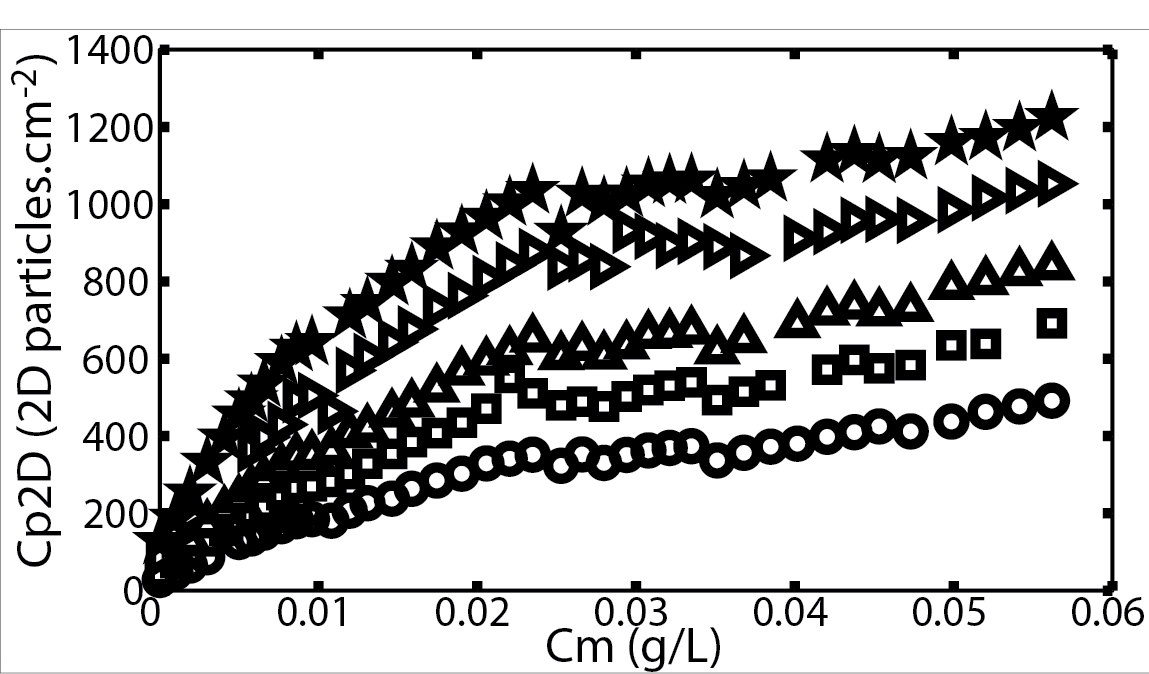}& 
\includegraphics[width=0.5\textwidth]{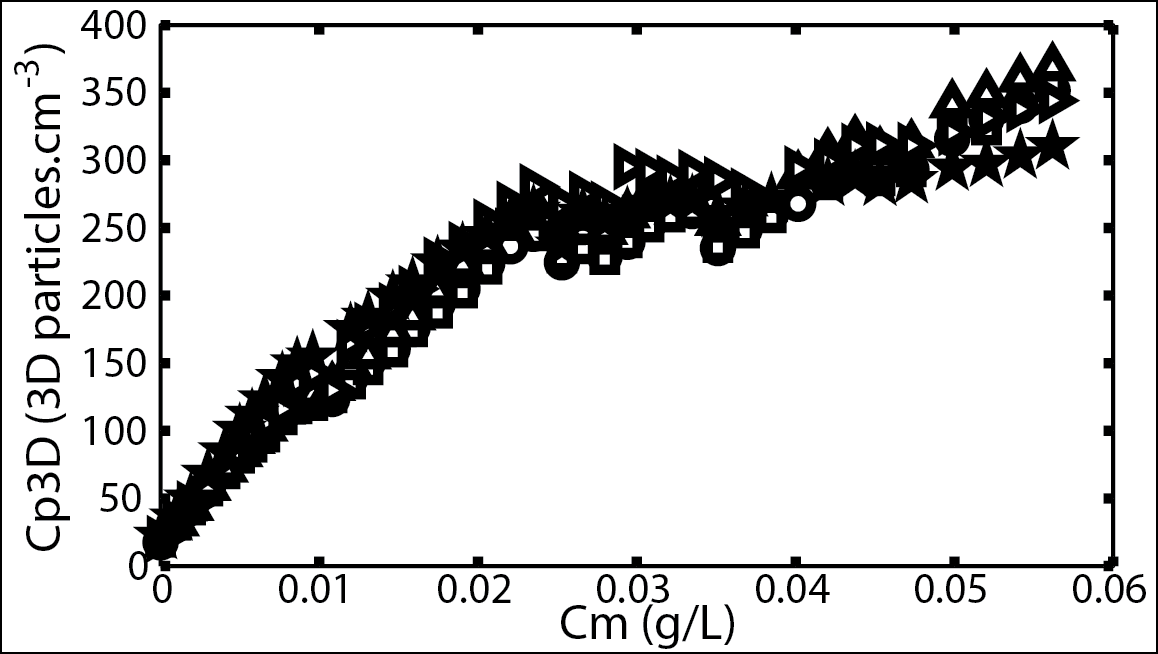} \\
a)&b)\\
\includegraphics[width=0.5\textwidth]{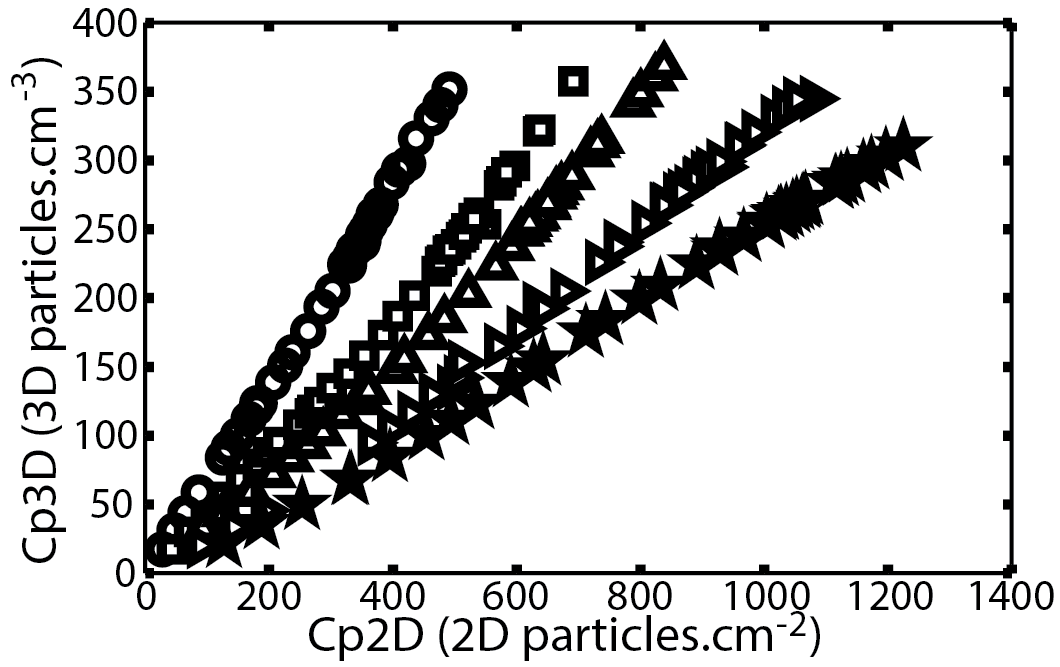}  & 
\includegraphics[width=0.5\textwidth]{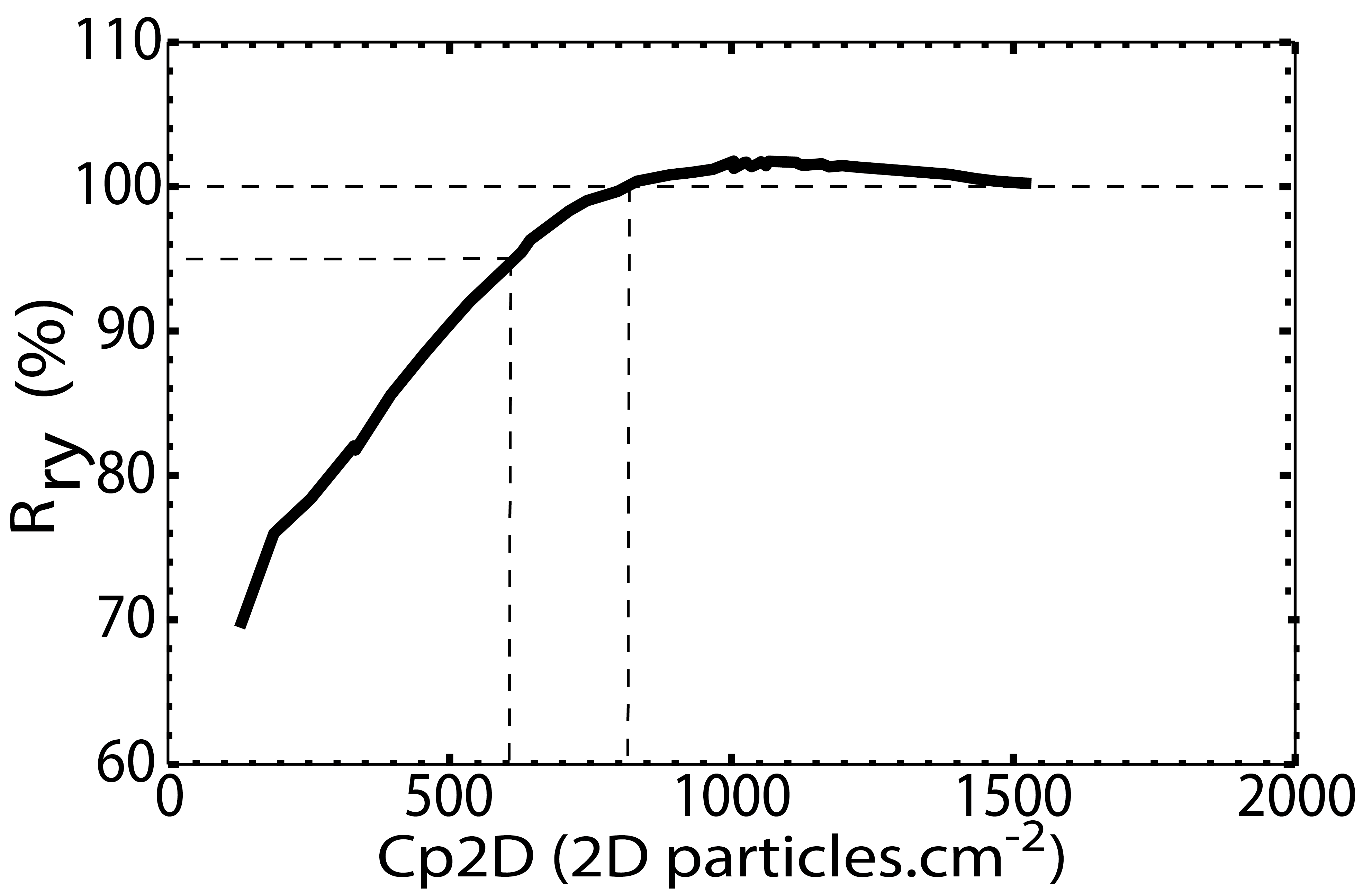} \\
c)&d)
\end{tabular}
\caption{Illumination thickness:  $h=$ 1cm ($\bigcirc$), $h=$ 1.5 cm ($\square$), $h=$ 2cm ($\triangle$), $h=$ 3cm ($\vartriangleright$), $h=$ 4cm ($\star$). a) Variation of the imaged concentration of 2D particles ($\mathrm{Cp_{2D}}$) with the mass concentration $\mathrm{C_{m}}$.
b) Variation of the concentration of 3D particles in the measurement volume ($\mathrm{Cp_{3D}}$) with $\mathrm{C_{m}}$.
c) Relationship between $\mathrm{Cp_{3D}}$ and $\mathrm{Cp_{2D}}$.
d) For $h=$4 cm, evolution of the efficiency of the reconstruction step (raw yield $\mathrm{R_{ry}}$ ) with the $\mathrm{Cp_{2D}}$.  }\label{Fig11}
\end{center}\end{figure}

The Fig. \ref{Fig11}a and \ref{Fig11}b present the variations of the particle concentrations ($\mathrm{Cp_{2D}}$ and $\mathrm{Cp_{3D}}$) with the mass concentration $\mathrm{C_{m}}$. As expected, the $\mathrm{Cp_{2D}}$ increases with the illumination thickness $h$ (Fig. \ref{Fig11}a). For low concentrations, the $\mathrm{Cp_{2D}}$ linearly follows the increase of mass concentration $\mathrm{C_{m}}$. For high concentrations, this is no longer the case because the images becomes saturated by the particles.
The $\mathrm{Cp_{3D}}$ does not depend on the illumination thickness $h$ (Fig. \ref{Fig11}b). 

The Fig. \ref{Fig11}c presents the evolution of $\mathrm{Cp_{3D}}$ as a function of $\mathrm{Cp_{2D}}$: for a given $\mathrm{Cp_{2D}}$, the thicker the measurement volume, the lower the $\mathrm{Cp_{3D}}$. 
 The dependence between these two variables being linear, one could think that the reconstruction step is not affected by the screening phenomenon. Actually, its effect on the reconstruction step is more subtle. Fig. \ref{Fig11}d presents the efficiency of the 3D particle reconstruction step (raw yield $\mathrm{R_{ry}}$) as a function of $\mathrm{Cp_{2D}}$. One observes that the raw yield becomes higher than 100 $\%$ for $\mathrm{Cp_{2D}} > 800$ particles.cm$^{-2}$. If more than 100 $\%$ of the detected particles are used during the 3D particles reconstruction step, it means some particles are used to rebuild two different triplets which should not be possible (ghost particles are disabled). 
The setting of the visual saturation in particle images has clearly an influence on the quality of the measurements. It points out the limitations of the reconstruction algorithms and it is currently not possible to detect this phenomenon properly.

\begin{figure}[h!tb]
\begin{center}
\begin{tabular}{cc}
\includegraphics[width=0.5\textwidth]{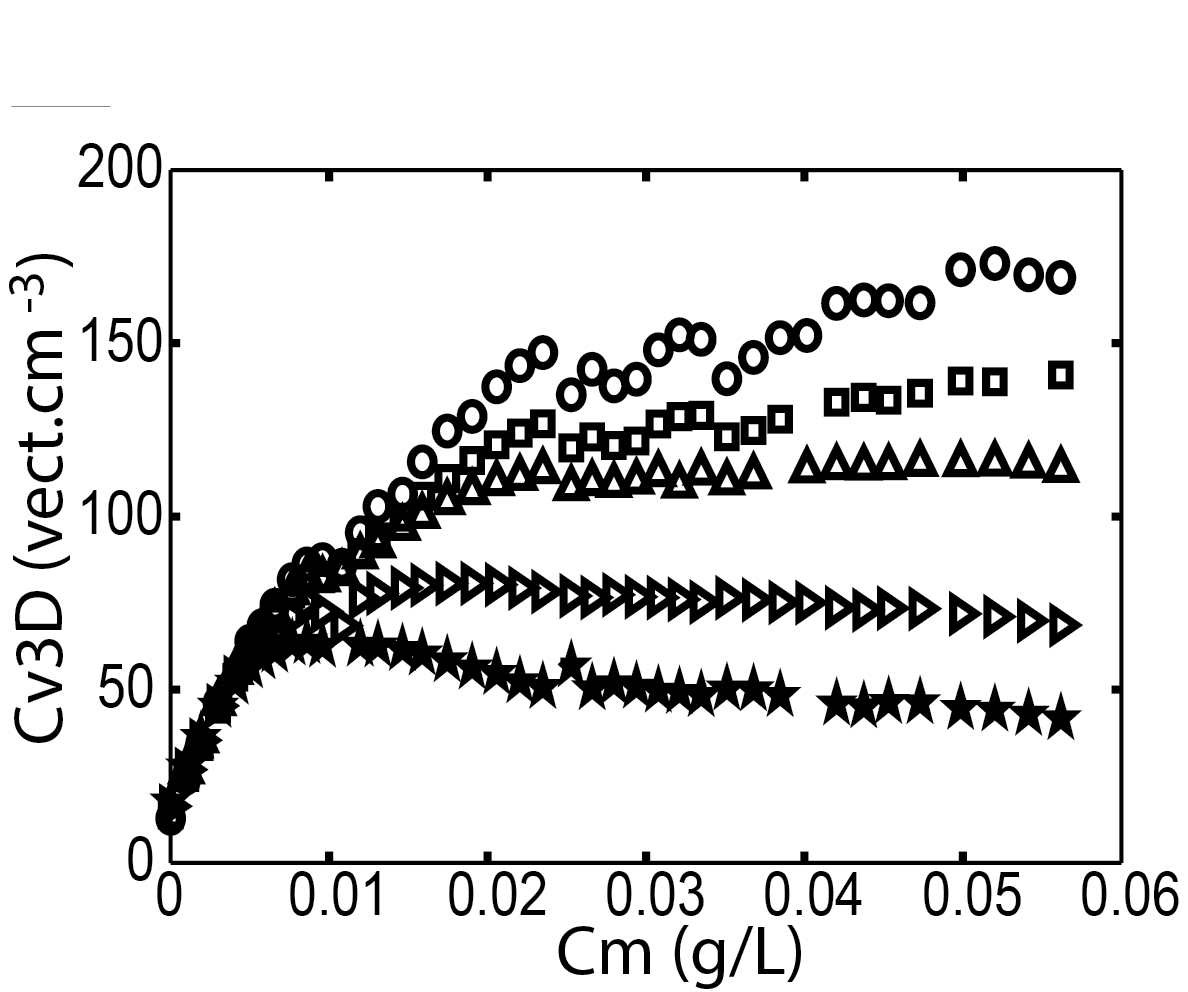}&
\includegraphics[width=0.49\textwidth]{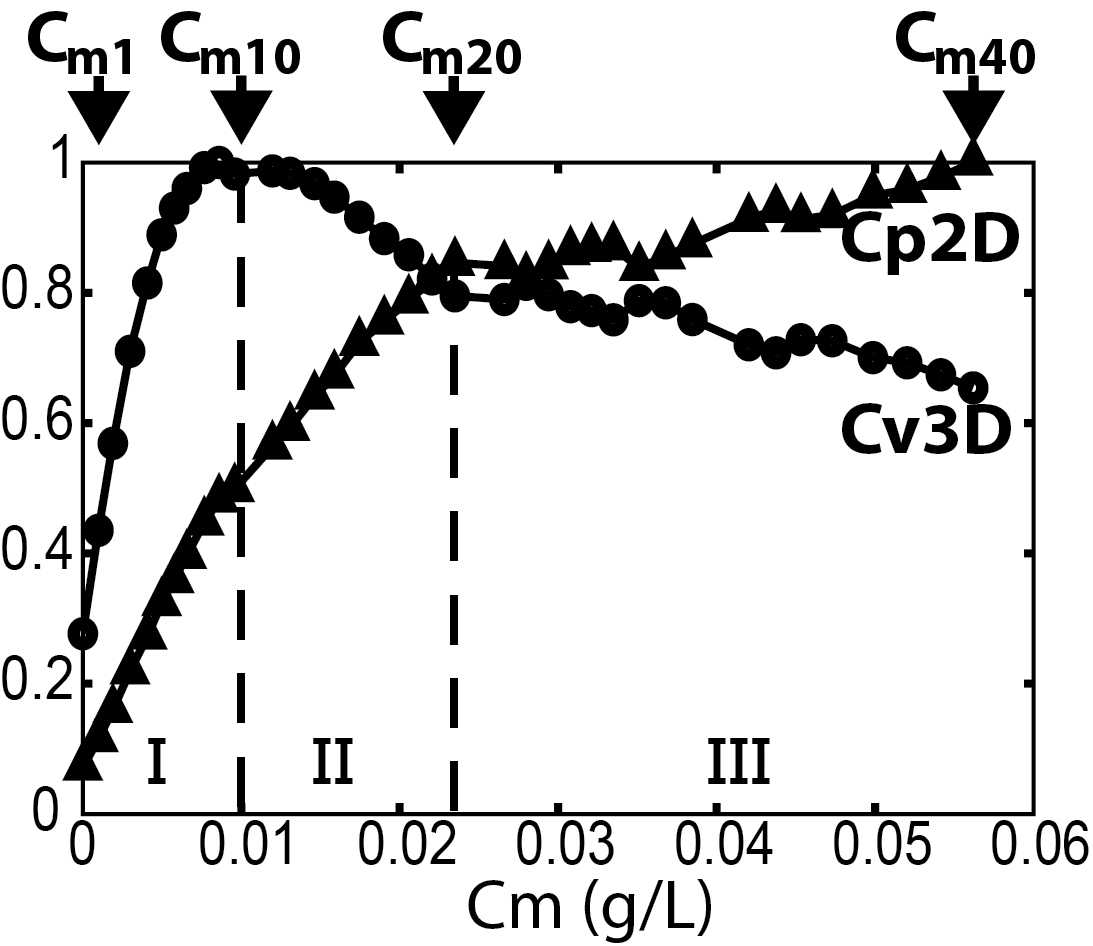} \\ 
a)&b)
\end{tabular}
\caption{a) Relationship between $\mathrm{Cv_{3D}}$ and $\mathrm{C_{m}}$.
Illumination thickness:  $h=1$ cm ($\bigcirc$),  $h=1.5$ cm ($\square$),  $h=2$ cm ($\triangle$), $h= 3$ cm ($\vartriangleright$), $h= 4$ cm ($\star$). b) For $h= 4$ cm,  evolution of $\mathrm{Cv_{3D}}$ ($\bigcirc$) and $\mathrm{Cp_{2D}}$  ($\square$) normalized by their maximal value, as a function of the mass concentration $\mathrm{C_{m}}$.  }\label{Fig12}
\end{center}\end{figure}

The Fig. \ref{Fig12}a presents the evolution of $\mathrm{Cv_{3D}}$ as a function of $\mathrm{C_m}$ for various illumination thicknesses.  For $h=3$ cm and $h=4$ cm, the concentration reaches a maximum which is followed by a slow decrease for higher $\mathrm{C_m}$. For these illumination thicknesses, the decrease of the $\mathrm{Cv_{3D}}$ can be attributed to lower experimental efficiency, the beginning of a statistically significant screening effect, as well as lower performances of the algorithms in the detection, reconstruction and tracking steps. The optical screening phenomenon is independent of the velocimetry system. On the opposite, the lower performances of the algorithms are associated with the intrinsic characteristics of the V3V system. 

The Fig. \ref{Fig12}b shows, for $h=4$ cm, the evolution of $\mathrm{Cp_{2D}}$ and $\mathrm{Cv_{3D}}$ (normalized by their maximum value to make comparison easier) as a function of $\mathrm{C_m}$.
$\mathrm{Cv_{3D}}$ is maximum for $\mathrm{C_{m}}=\mathbf{C_{m10}}=0.01$ g.L$^{-1}$. For the same mass concentration, a small inflection of the $\mathrm{Cp_{2D}}$ curve is observed.  A stronger inflection of $\mathrm{Cp_{2D}}$  occurs around $\mathrm{C_{m}}=\mathbf{C_{m20}}=0.023$ g.L$^{-1}$, which also corresponds to a clear change of the slope of the  $\mathrm{Cv_{3D}}$ curve. The concentration $\mathrm{C_{m10}}$ and $\mathrm{C_{m20}}$ correspond to these two successive inflections points and define the boundaries of three different regimes for $\mathrm{Cp_{2D}}$ and $\mathrm{Cv_{3D}}$. 

\subsection{Influence of $\mathrm{C_{m}}$ on the 2D and 3D reconstruction steps}
Depending on the mass concentration, one can distinguish three main regimes for the 3D and 2D particle concentrations on Fig. \ref{Fig12}b:

\begin{itemize}
\item \textbf{Regime I:} For $\mathrm{C_{m1}} < \mathrm{C_{m}} < \mathrm{C_{m10}}$ (Fig. \ref{Fig9}\emph{b} and \ref{Fig9}\emph{f}), the inter-particles mean distance remains significantly larger than the average diameter of the detected particles. The 2D particle concentration remains sufficiently low to efficiently carry out the identification, reconstruction and tracking steps. Therefore, increasing the mass concentration of particles directly leads to an increase of $\mathrm{Cp_{2D}}$ and $\mathrm{Cv_{3D}}$. In this regime, the concentration are well-suited for the V3V system.

\item \textbf{Regime II:} For $\mathrm{C_{m10}} < \mathrm{C_{m}} < \mathrm{C_{m20}}$, the concentration of particles seen by the cameras increases. The Fig. \ref{Fig9}\emph{c} and \ref{Fig9}\emph{g} show the formation of aggregates resulting from the visual overlapping of particle images. Even if it can be considered as an early stage of the optical screening phenomenon, the background is still visible and 
most of the particles can still be detected during the 2D particle identification step, and the 3D particle reconstruction step can still be completed. Nevertheless the tracking algorithm reaches its limits. Beyond a critical 3D particle concentration, it becomes more and more difficult to identify the target particle at $t+\delta t$, lowering the efficiency of the velocity processing step. 
The decrease of  $\mathrm{Cv_{3D}}$  seen in the Fig. \ref{Fig12}b is not governed by the screening effect, but by a decreasing efficiency of the tracking algorithms. 
  For this reason, this regime is called \emph{algorithmic decay} and is specific to the measurement system.  The critical concentration for which algorithmic saturation begins is then $C_{m\ alg.sat}=C_{m10}$. 
\item \textbf{Regime III : }  Between $\mathrm{C_{m20}}$ and $\mathrm{C_{m40}}$ (Fig. \ref{Fig9}\emph{d} and \ref{Fig9}\emph{h}), the images becomes saturated both in particles and in pixel intensity. The mean diameter of the detected particles has considerably increased compared to  $C_{m1}$. The critical concentration for which screening effects become predominant is the $C_{m\ opt.scr}=C_{m20}$.
\end{itemize}

\begin{figure}[htb!]
 \begin{center} 
 \begin{tabular}{cc}
\includegraphics[width=0.51\textwidth]{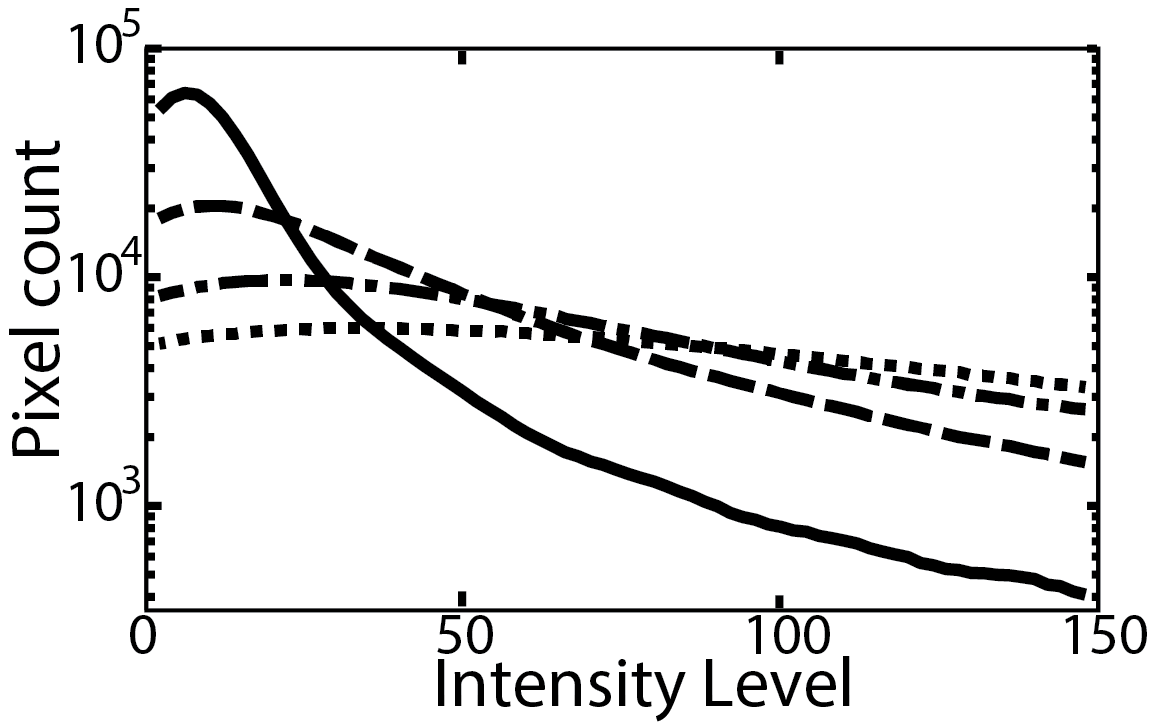}& 
\includegraphics[width=0.48\textwidth]{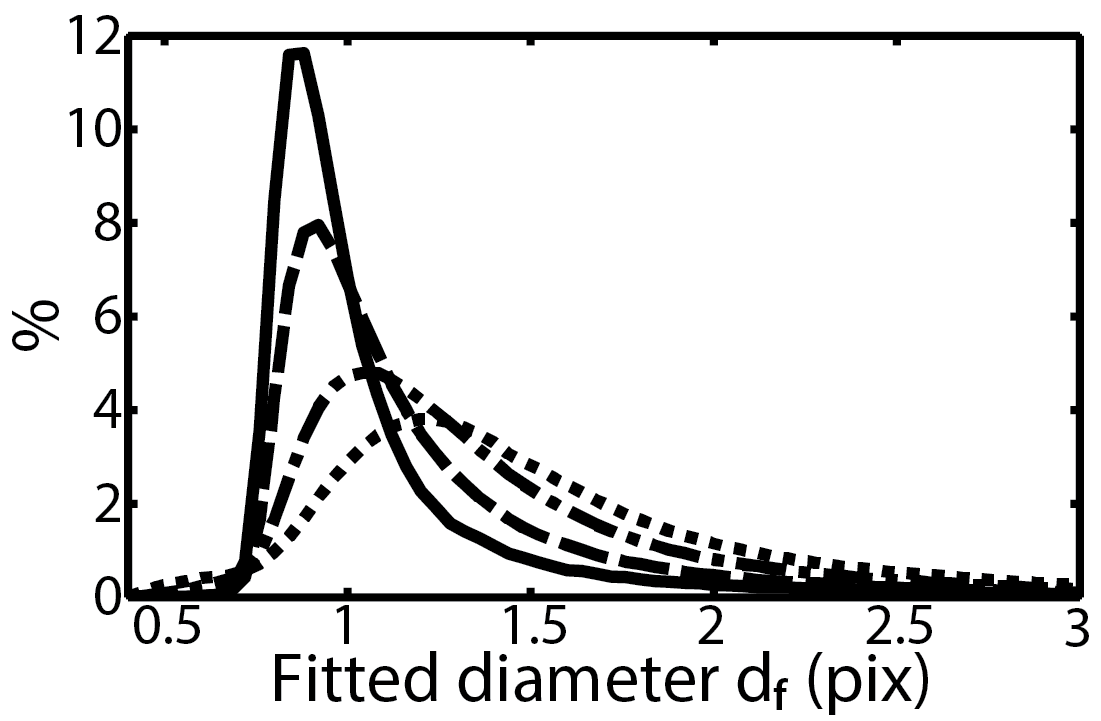} \\
a)&b)
\end{tabular}
\caption{a) Illumination thickness $h=4$ cm, $PDT = 30$. Pixel count of a given intensity in instantaneous snapshots. The intensity is scaled between 0 and 4095. b) Gaussian-fitted diameter distribution of 2D detected particles for different seeding mass concentrations.
Mass concentrations: $-$ $C_{m1}$, $---$ $C_{m10}$, $- \cdot - $ $C_{m20}$, $\cdots$ $C_{m40}$.}\label{Fig13}
\end{center}\end{figure}

It is illustrated in the Fig. \ref{Fig13}a and \ref{Fig13}b.\\ 
The Fig.\ref{Fig13}a shows in semilog the intensity distribution of a given image. The intensity is discretized over 12 bits. For the sake of clarity, only the first 150 values are represented. The progressive homogenization of the image intensity distribution happens in two steps:

\begin{itemize}
\item First of all, for $C_{m1} \leq C_m \leq C_{m20}$, the evolution of the intensity distribution is dominated by a flattening of the distribution of the pixel intensity  which corresponds to the increasing number of particles in the images. When $C_m = C_{m20}$, the balance of intensities is considerably more uniform. Nevertheless the intensities of the pixels of the background ($\mathrm{intensity\ level}<12\sim15$) are still four times higher than the typical particle intensities. Therefore, the background intensity still prevails. 
\item For $\mathrm{C_m > C_{m20}}$, the distribution of the pixel intensities becomes more and more homogenous.
As a result, the particles are no longer bright spots over a dark background. Instead, they rather lay over a bed of moving particles. It corresponds to the optical screening of the particles by each others. The average intensity of the background increases and disturbs strongly the particle detection. The intensity peaks associated with the centers of the particles are drowned in the background which makes impossible the detection of the less luminous particles. Despite the background removal preprocessing, the visual saturation of particle images prevent a good detection of the 2D particles as well as the reconstruction of the 3D particles.

\end{itemize}

As a consequence, the distributions of diameters of the detected particles  for the different concentrations is also modified, as illustrated on Fig.\ref{Fig13}b: 

\begin{itemize}
\item For $\mathrm{C_{m} < C_{m20}}$ (Fig. \ref{Fig9}\emph{a, b, e, f}), the peak of the distribution of the diameters is well defined around $d_f \approx$ 1 pixel. 
\item For $\mathrm{C_{m} > C_{m20}}$ (Fig. \ref{Fig9}\emph{c, d, g, h}), the diameter distribution widens more and more and the position of the peak shifts toward increasing diameters. Whereas the real size of the particles do not change, the increase of the estimated average diameter shows the limitation of the detection algorithms. Due to the particle overlapping phenomenon, the algorithms can no longer determine the particle\rq{}s boundaries and no longer measure properly the particle diameter as the gaussian fit includes several particles. This leads to a degradation of the performances of the detection and reconstruction algorithms. 
\end{itemize}

\subsection{Relation between $\mathrm{Cv_{3D}}$ and $R_{\% cover}$}

The evolutions of $\mathrm{Cv_{3D}}$ and $R_{\% cover}$ as a function of $\mathrm{Cp_{2D}}$ have been investigated for various illumination thicknesses and voxel sizes.

\begin{figure}[htb!]
\begin{center}
\begin{tabular}{cc}
\includegraphics[width=0.49\textwidth]{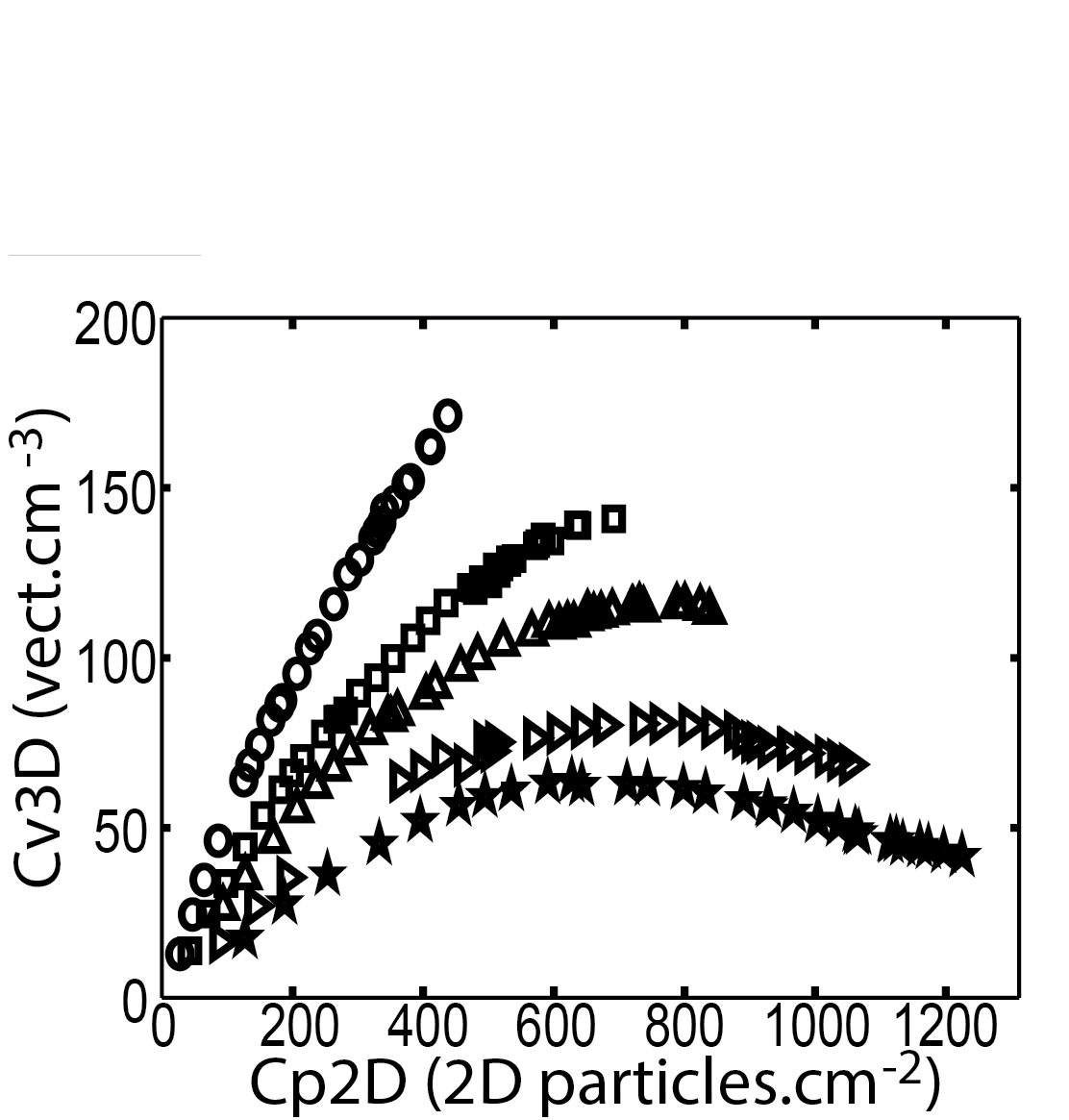}& 
\includegraphics[width=0.52\textwidth]{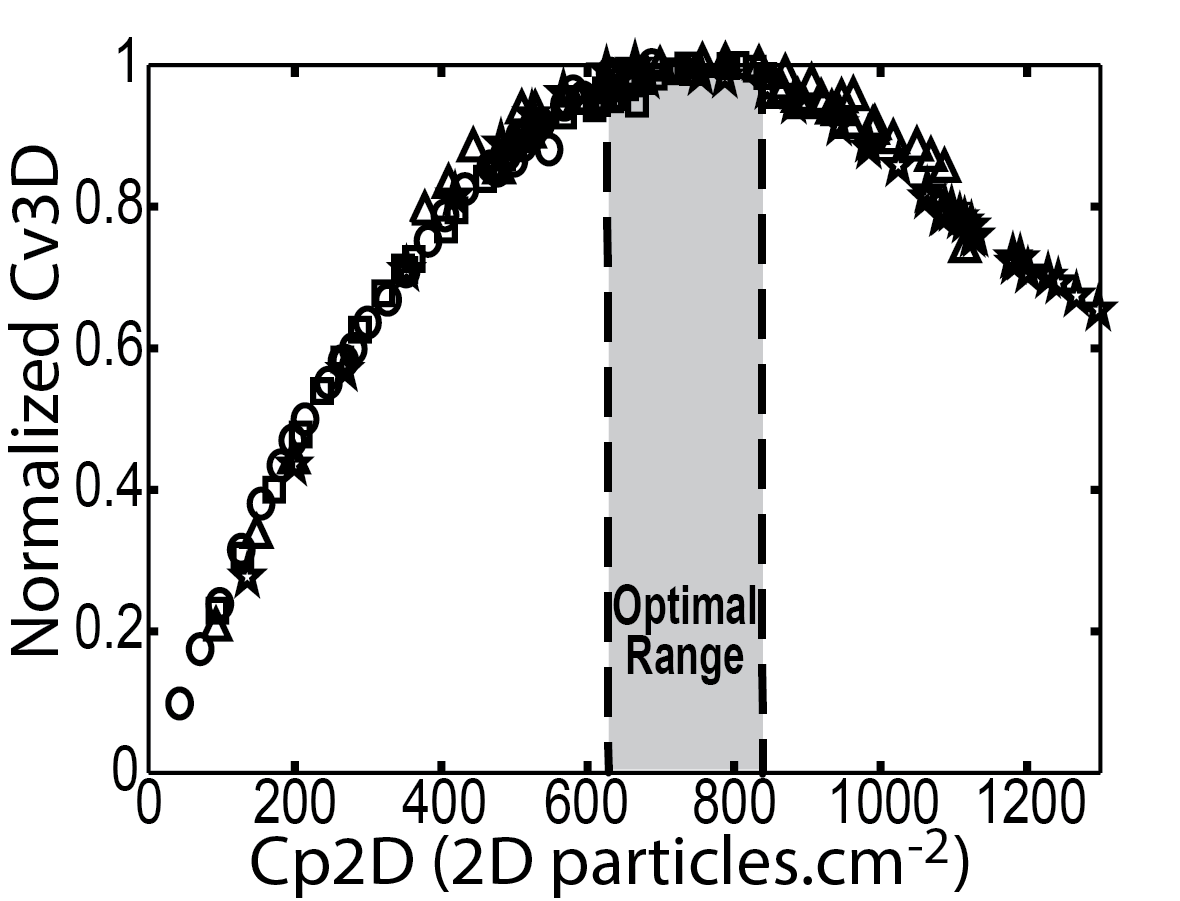}\\
a)&b)
\end{tabular}
\caption{a) $\mathrm{Cv_{3D}}$ as a function of $\mathrm{Cp_{2D}}$. b) $\mathrm{Cv_{3D}}$ normalized by the maximum of each curve as a function of $\mathrm{Cp_{2D}}$. Illumination thickness: $h=$ 1cm ($\bigcirc$), $h=1.5$ cm ($\square$),  $h= 2$ cm ($\triangle$), $h = 3$ cm ($\vartriangleright$), $h= 4$ cm ($\star$).}
\label{Fig14}
\end{center}
\end{figure}

In the Fig. \ref{Fig14}a, $\mathrm{Cv_{3D}}$ is maximum for a $\mathrm{Cp_{2D}}\approx$ 730 part.cm$^{-2}$, whatever the illumination thickness. It becomes obvious when $\mathrm{Cv_{3D}}$ is normalized by the maximum of each curve (Fig. \ref{Fig14}b). 
$\mathrm{Cp_{2D}}$ values for $h=3$ cm and $h=4$ cm reach and exceed the threshold value of 730 part.cm$^{-2}$. One can define a range of $\mathrm{Cp_{2D}}$ optimizing $\mathrm{Cv_{3D}}$ (620 $< \mathrm{Cp_{2D}} <$ 840 particles.cm$^{-2}$) and an optimal concentration $(\mathrm{Cp_{2D}} \approx$ 730 particles.cm$^{-2}$).
 Taking that into account, we use the $\mu m$ to pixel conversion factors displayed in the table \ref{tableLpxlocal} to provide absolute concentrations in ppp. As shown in table \ref{tableLpxlocal}, the relation between this conversion factor and the distance along the optical axis is linear. Therefore the conversion factor of a given measurement volume can be computed as the mean value of the conversion factor  of the front and back planes. This leads to an optimal $\mathrm{Cp_{2D}}$ range: $3.8\times 10^{-2}<Cp_{2D}<5.2\times 10^{-2}$ ppp centered around $\mathrm{Cp_{2D}}\approx 4.5\times 10^{-2}$ ppp.  
These 2D concentration values are relative to our experiments, since they depend on the camera location and the typical particle size. Nevertheless the camera location and the typical particle size have also been chosen to be provide optimal measurements, these $\mathrm{Cp_{2D}}$ values can be taken as representative of the optimal achievable performances of the V3V system.\\

\begin{figure}[htb!]
\begin{center}
\begin{tabular}{cc}
\includegraphics[width=0.49\textwidth]{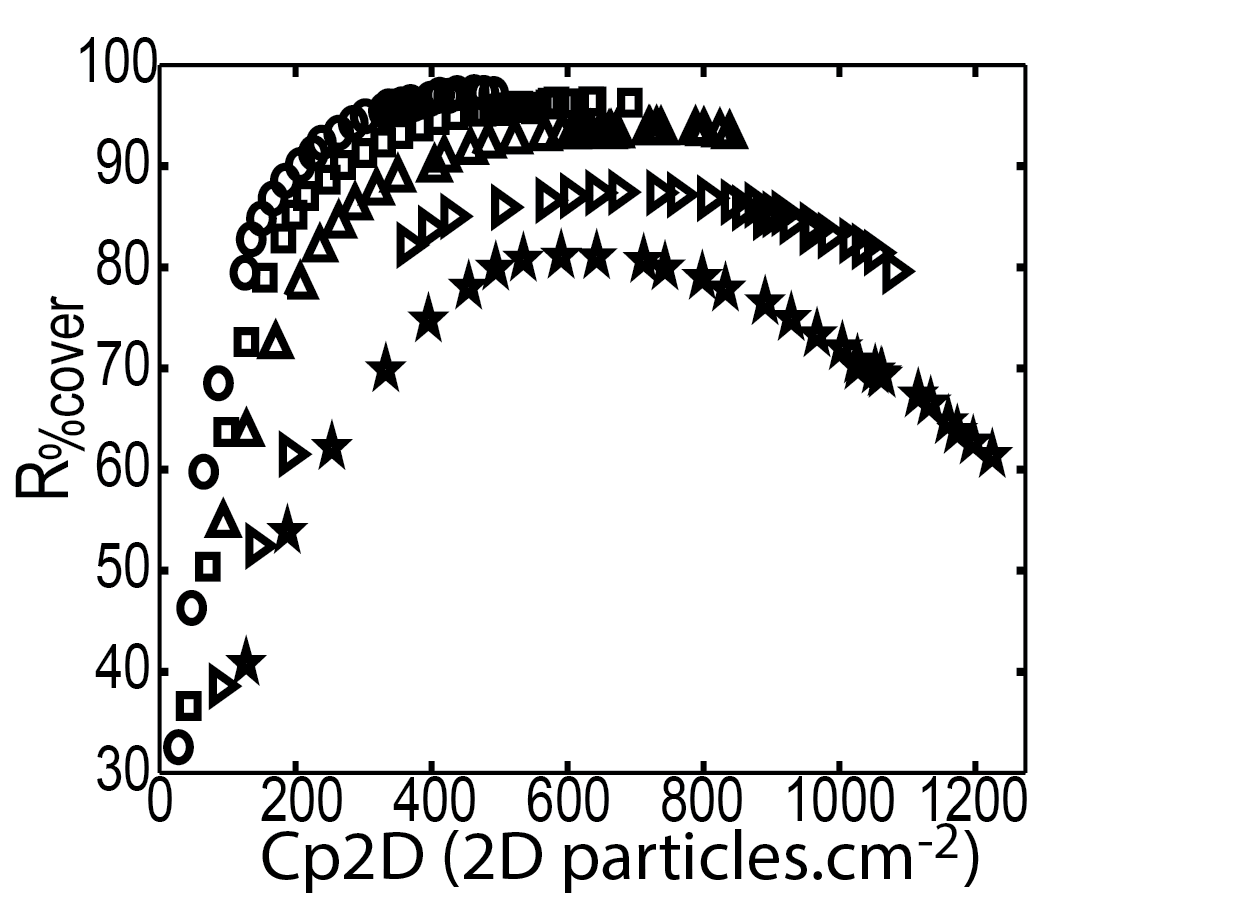}&
\includegraphics[width=0.5\textwidth]{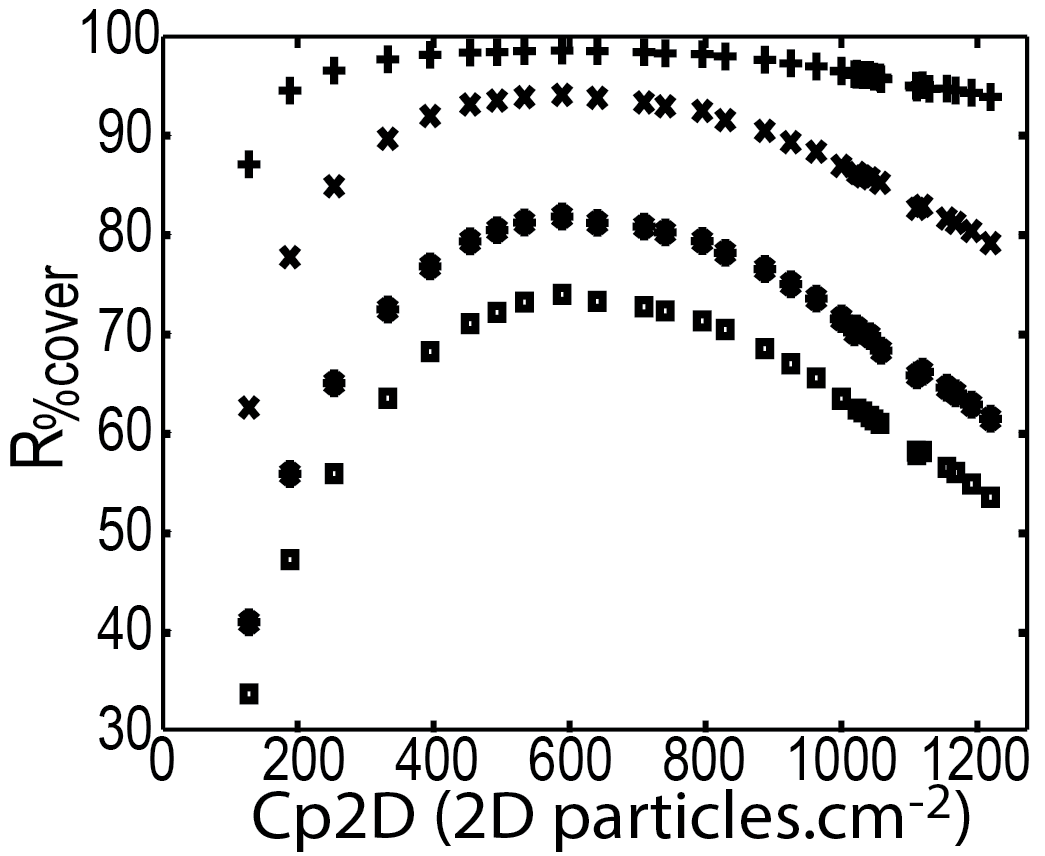}\\
a)&b)
\end{tabular}
\caption{a) Voxel size = 0.24 cm. Covering percentage as a function of $\mathrm{Cp_{2D}}$.
Illumination thickness:  $h=$1 cm ($\bigcirc$), $h=$1.5 cm ($\square$),  $h=$ 2cm ($\triangle$), $h=$ 3cm ($\vartriangleright$), $h=$ 4cm ($\star$).  b) $h=$ 4 cm. covering percentage according to $\mathrm{Cp_{2D}}$ for different voxel sizes.
Voxel size: $\mathrm{l_{vox}}=$ 0.4 cm (+), $\mathrm{l_{vox}}=$0.3 cm (x),  $\mathrm{l_{vox}}=$0.24 cm ($\star$), $\mathrm{l_{vox}}=$0.2 cm ($\square$).}
\label{Fig15}
\end{center}
\end{figure}

The Fig. \ref{Fig15}a shows the evolution of the covering percentage $\mathrm{R_{\%cover}}$ as a function of $\mathrm{Cp_{2D}}$ for various illumination thicknesses and for a given voxel size $\mathrm{l_{vox}}=0.24$ cm. One can see that a maximum is clearly reached for $h=3$ cm and $h=4$ cm. The optimal $\mathrm{Cp_{2D}}$ is independent of the illumination thickness. The illumination thickness only influences the value of the maximum of $\mathrm{R_{\%cover}}$. Similar results have been obtained for different voxel sizes but are not presented.
 \begin{figure}[htb!]
\begin{tabular}{cc}
\includegraphics[width=0.49\textwidth]{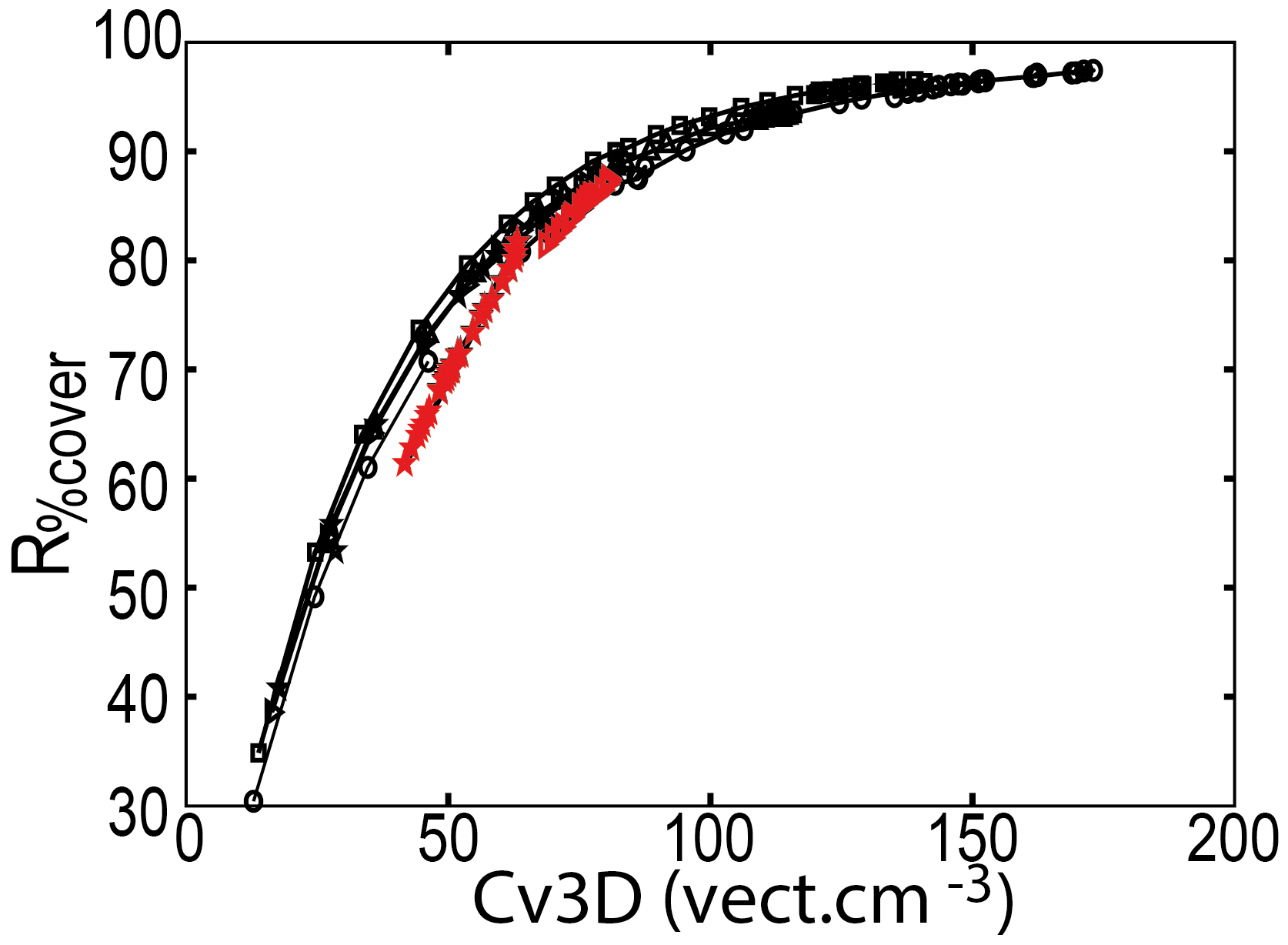}& 
\includegraphics[width=0.49\textwidth]{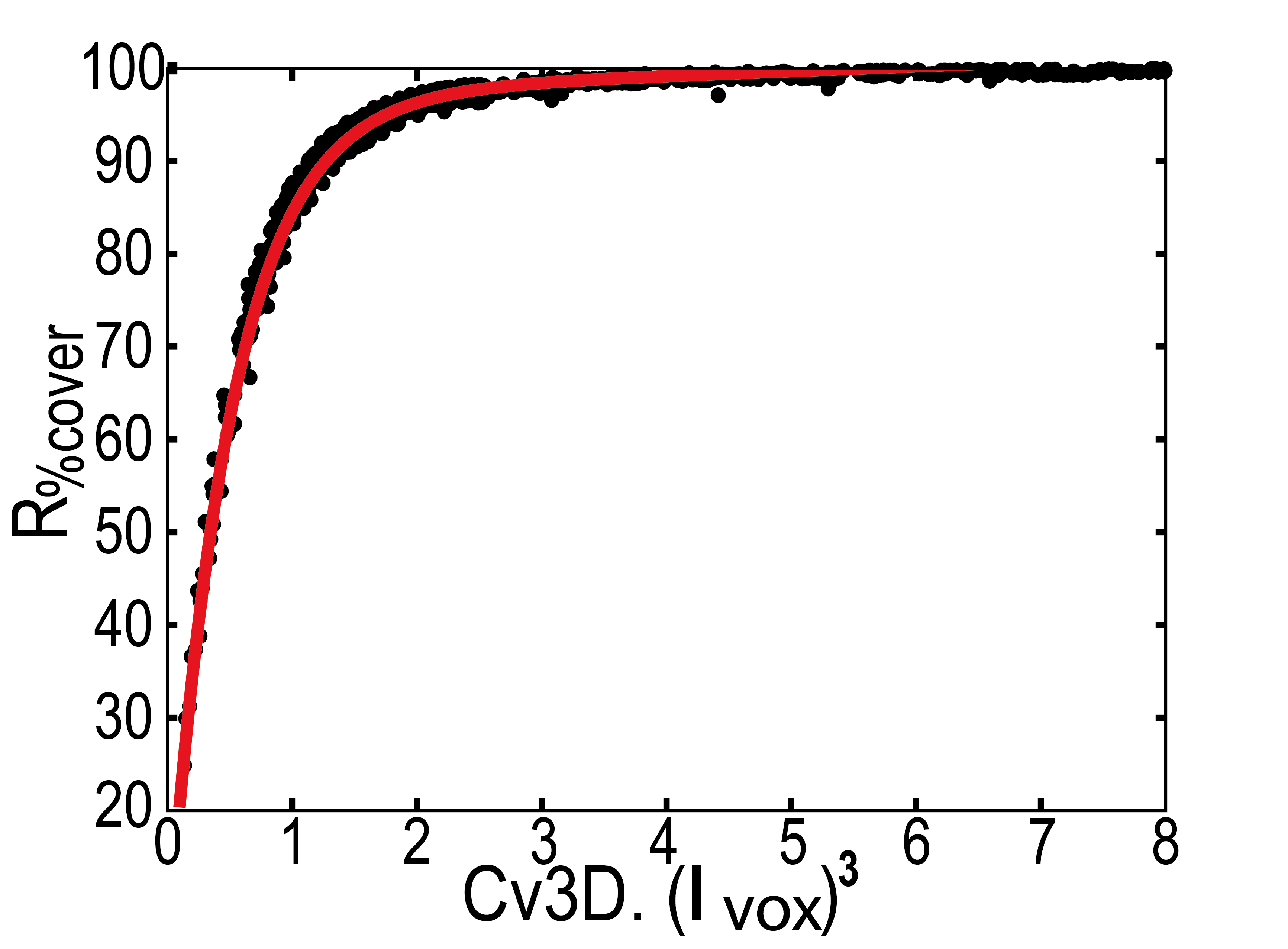}\\
a)&b)
\end{tabular}
\begin{center}
\end{center}
\caption{a) $\mathrm{\mathrm{R_{\%cover}}}$ as a function of $\mathrm{Cv_{3D}}$. $\mathrm{l_{vox}}=0.24$cm. b)  For all illumination thicknesses $h=\{ 1, 1.5, 2, 3, 4\}$ cm, and all voxel sizes $\mathrm{l_{vox}}=$ \{0.8, 0.7, 0.6, 0.5, 0.4, 0.36, 0.32, 0.3, 0.28, 0.26, 0.24, 0.22, 0.2\} cm, $\mathrm{R_{\%cover}}$ as a function of $\mathrm{Cv_{3D}}$ scaled by $\mathrm{l_{vox}}^3$. Illumination thickness: $h=$ 1cm ($\bigcirc$), $h=$1.5 cm ($\square$),  $h=$2 cm ($\triangle$),  $h=$3 cm ($\vartriangleright$), $h=$4 cm ($\star$).}
\label{Fig16}
\end{figure}

The Fig. \ref{Fig15}b presents the evolution of $\mathrm{R_{\%cover}}$ as a function of $\mathrm{Cp_{2D}}$ for different values of the voxel size and for $h=$ 4 cm. It confirms that for a given illumination thickness, the position of this maximum is independent of the voxel size, i.e. of the interpolation step. Similar curves for different voxel sizes are not presented but show the same trend.

The Fig. \ref{Fig16}a shows that  the evolution of $\mathrm{R_{\%cover}}$ as a function of $\mathrm{Cv_{3D}}$ (for a voxel size $\mathrm{l_{vox}}=0.24$ cm) does not depend on the illumination thicknesses. The relation between $\mathrm{R_{\%cover}}$ and $\mathrm{Cv_{3D}}$ and the efficiency of the interpolation step is then entirely characterized by a single curve. A cusp (red markers in Fig. \ref{Fig16}a) can be seen on the h=3 and 4 cm evolutions. It corresponds to the optical screening and algorithmic decay regimes: $\mathrm{Cv_{3D}}$ and $\mathrm{R_{\%cover}}$ decrease after the optimal $\mathrm{Cp_{2D}}$.

The Fig. \ref{Fig16}b presents $\mathrm{R_{\%cover}}$ as a function of $\mathrm{Cv_{3D}}$ for different illumination thicknesses and also for different voxel sizes. If $\mathrm{Cv_{3D}}$ is scaled by $\mathrm{l_{vox}}^3$ (which is not the final grid resolution, but the size of the interpolation voxels), one can see that all the experimental curves collapse. An empirical fitting law is proposed: 
\begin{equation}
\mathrm{\mathrm{R_{\%cover}}}=\alpha_1\cdot\exp(\beta_1\cdot \mathrm{Cv_{3D}}\cdot \mathrm{l_{vox}}^3)-\alpha_2\cdot\exp(-\beta_2\cdot \mathrm{Cv_{3D}}\cdot \mathrm{l_{vox}}^3)\label{equationcouvertureCv3D}
\end{equation}
where $\alpha_1 =$ 96.66, $\beta_1 = 0.05$, $\alpha_2 =$ 93.35 and $\beta_2 =$ 2. This relation corresponds to the best fit of the experimental data. Thanks to this relation, it is possible to estimate the final covering percentage  when $\mathrm{Cv_{3D}}$ and $\mathrm{l_{vox}}$ are known, or on the contrary to estimate the necessary raw velocity vector concentration (and therefore the imaged concentration and maximal illumination thickness of the measurement volume) for given values of the final covering percentage and the voxel size $\mathrm{l_{vox}}$.
This relation can also be helpful to choose the experimental parameters according to the goals of the experiment.

Because parameters such as illumination thickness and final interpolation grid resolution are indirectly linked through the different processing concentrations (Fig. \ref{Fig7}), a balance has to be found between the measurement volume size and the final resolution of interpolated vectors. It strongly influences the resolution or the typical size of the flow structures which can be observed in the instantaneous velocity fields.  

As clearly demonstrated by \citet{Kahler2012}, PTV based systems potentially lead to the best possible resolution with pixel based cameras: it is limited by the precision of the Gaussian-fit algorithm which evaluates the sub-pixel coordinates of the particle center position. Nowadays, a  sub-pixel precision of 0.1 pixel is achievable in the best scenario. As mentioned in section \ref{sec:Processing},  typical values for the V3V system correspond to 0.2 pixel. It means the spatial resolution can be highly increased to match very fine resolution, unachievable with other velocimetry techniques \cite{Kahler2012}.

If this method is well indicated for classical time-averaging and phase-averaging processes, it can also be used in the case of quasi-periodic flows by using an appropriate conditional averaging (\citet{CAMBONIEthesis}). All relevant raw velocity fields can be united in a unique highly-dense raw velocity vectors field that can then be used as an input for the interpolation step.

\section{Application to a Jet in Cross Flow}

\subsection{Description of a jet in cross-flow}
This methodology is now applied to the case of a jet in crossflow (JICF), i.e. a transverse jets interacting with a boundary layer. This flow has been studied for several decades and continues to be an active subject of research for many experimental or numerical research teams.  Good reviews on the subject can be found in \citet{Karagozian2010,CamAi2014}. This complex three-dimensional and non stationary flow exhibits many vortical structures growing into a boundary layer. Characterizing such a complex 3D flow is still an experimental challenge and is an excellent test case to validate our methodology.

\subsection{Experimental parameters and methodology}

\begin{figure}
\centering
\centerline{\includegraphics[width=0.65\textwidth]{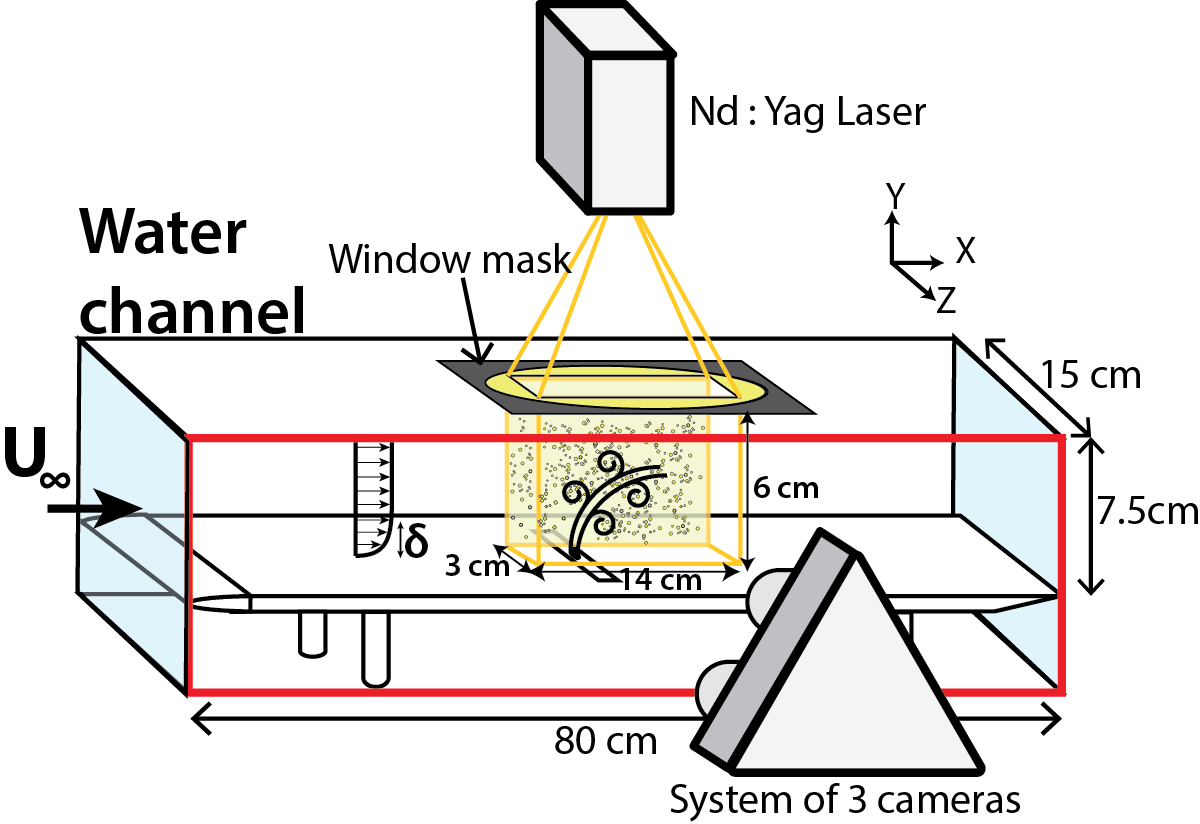}}
\caption{Sketch of the experimental setup for the jet in crossflow experiments.}
\label{Fig17}
\end{figure}

This experiment was carried out in the same hydrodynamic channel. The diameter of the circular nozzle of the jet is $d_{jet} = 0.8$ cm. It is flush-mounted on a flat plate, and the jet exit is located 40 cm downstream the leading edge of the flat plate in the symmetry plane. Compared to the previous experiments, the three cameras have been fixed along the side of the channel to optimize the resolution in the (X,Y) plane (Fig. \ref{Fig17}). We keep the Z axis along the optical axis of the system of cameras and the origin of the measurement volume is located at the center of of the exit of the circular nozzle.
The crossflow velocity is $U_{\infty} = 6.4\ $cm.s$^{-1}$ and the jet velocity is $V_{j} = 8\ $cm.s$^{-1}$, leading to a velocity ratio  $VR = V_{j}/U_{\infty} = 1.25$. The boundary layer is still laminar and its thickness just upstream the jet is $\delta_{99} = 1.39\ $cm. The crossflow Reynolds number is $Re_{cf} = U_{\infty}\cdot\delta_{99}/\nu \approx 890$, the jet Reynolds number $Re_{jet} = V_{jet}\cdot d_{jet}/\nu \approx 640$ and the Reynolds number commonly used for the jet in crossflow $Re_{JICF} = U_{\infty}\cdot d_{jet}/\nu \approx 512$.

The work of \citet{Gopalan2004} on round jets with low velocity ratio reveals that most of the JICF dynamic happens in a domain of approximately $\pm1.5 d_{jet}$ around the center of the round jet. 
For this experiment, the illumination volume thickness, centered on the exit of the jet,  is $h = 3\ $cm. 
Since optimal values of $\mathrm{Cp_{2D}}$ range from 620 to 840 part.cm$^{-2}$, a $\mathrm{Cp_{2D}}=$ 680 part.cm$^{-2}$ has been chosen for these experiments. It approximately corresponds to 51 000 2D particles in a region of interest which covers 74.85 cm$^{2}$ (particle processing mask area). For this illumination thickness, one can expect $Cv_{3D} \approx$ 75 vect.cm$^{-3}$ (Fig. \ref{Fig14}a). The objective is to at least obtain a 99$\%$ coverage of the instantaneous fields. \\
Using equation \ref{equationcouvertureCv3D}, $\mathrm{R_{\%cover}}=99\%$ corresponds to $Cv_{3D\ adim} = 4.48$. Since $Cv_{3D\ adim} = Cv_{3D}\cdot (\mathrm{l_{vox}}) ^{3}$, the voxel size is $\mathrm{l_{vox}}=\sqrt[3]{\frac{Cv_{3D\ adim}}{Cv_{3D}}}=$0.39 cm. The optimal voxel size for the last stage of interpolation is thus approximately 0.4 cm. Considering a 75 $\%$ overlapping of the voxels, it leads to a final velocity vector field with one vector every 0.1 cm.

\subsection{Instantaneous velocity fields}
\begin{figure}[htb!]
\begin{center}
 \begin{tabular}{cc}
\includegraphics[width=0.5\textwidth]{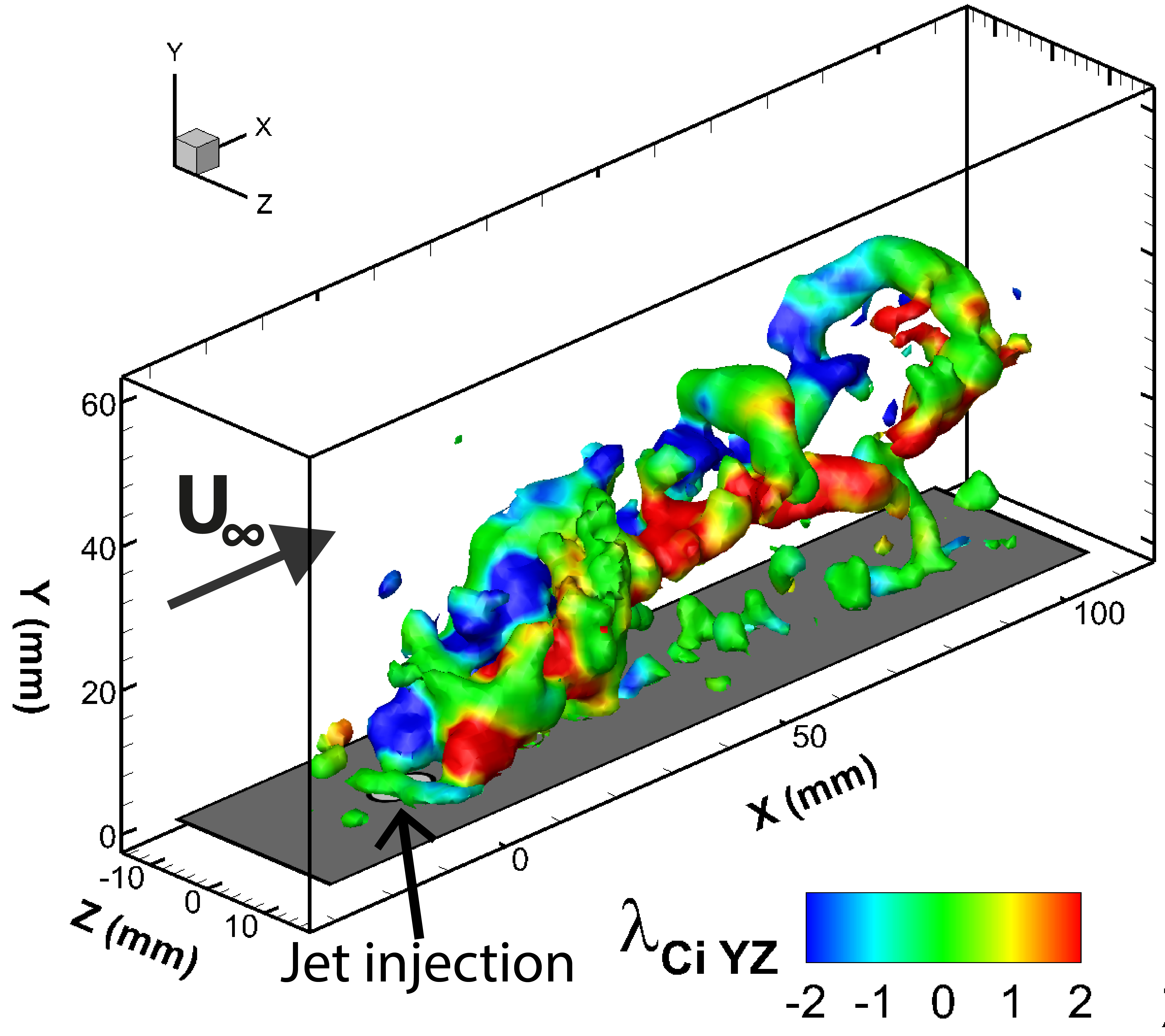} &
\includegraphics[width=0.5\textwidth]{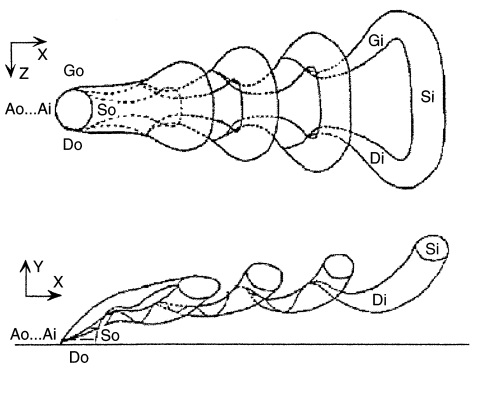}\\
a)&b)
\end{tabular}
\end{center} 
\caption{a) Visualization of the vortices in the instantaneous velocity field using isosurfaces of $\lambda_{Ci}$ colored in $\lambda_{Ci YZ}$ the longitudinal swirl in the( Y, Z) planes. b) Sketch of the  vortices expected for this velocity ratio as proposed by \citet{Blanchard1999}.}\label{Fig18}
\end{figure}

A visualization of an instantaneous velocity field is shown on Fig. \ref{Fig18}a using an isosurface of $\lambda _{Ci}$ \citep{Zhou1999,Christensen2002}. 

The hairpin vortical structures characteristic of the low velocity ratios is recovered, in agreement with the scenario suggested by \citet{Blanchard1999} (see Fig.\ref{Fig18}b).  
The optimization of the seeding and minimization of screening effects makes it possible to obtain very good \emph{instantaneous} velocity fields with a high spatial resolution for instantaneous volumetric velocimetry measurements. This condition is necessary to be able to visualize the 3D hairpin-like vortices which disappear in the time-averaged velocity field.

\subsection{Validation of the choice of the size of the voxels}
\begin{figure}[htb!]
 \begin{center}
 \begin{tabular}{cc}
\includegraphics[width=0.5\textwidth]{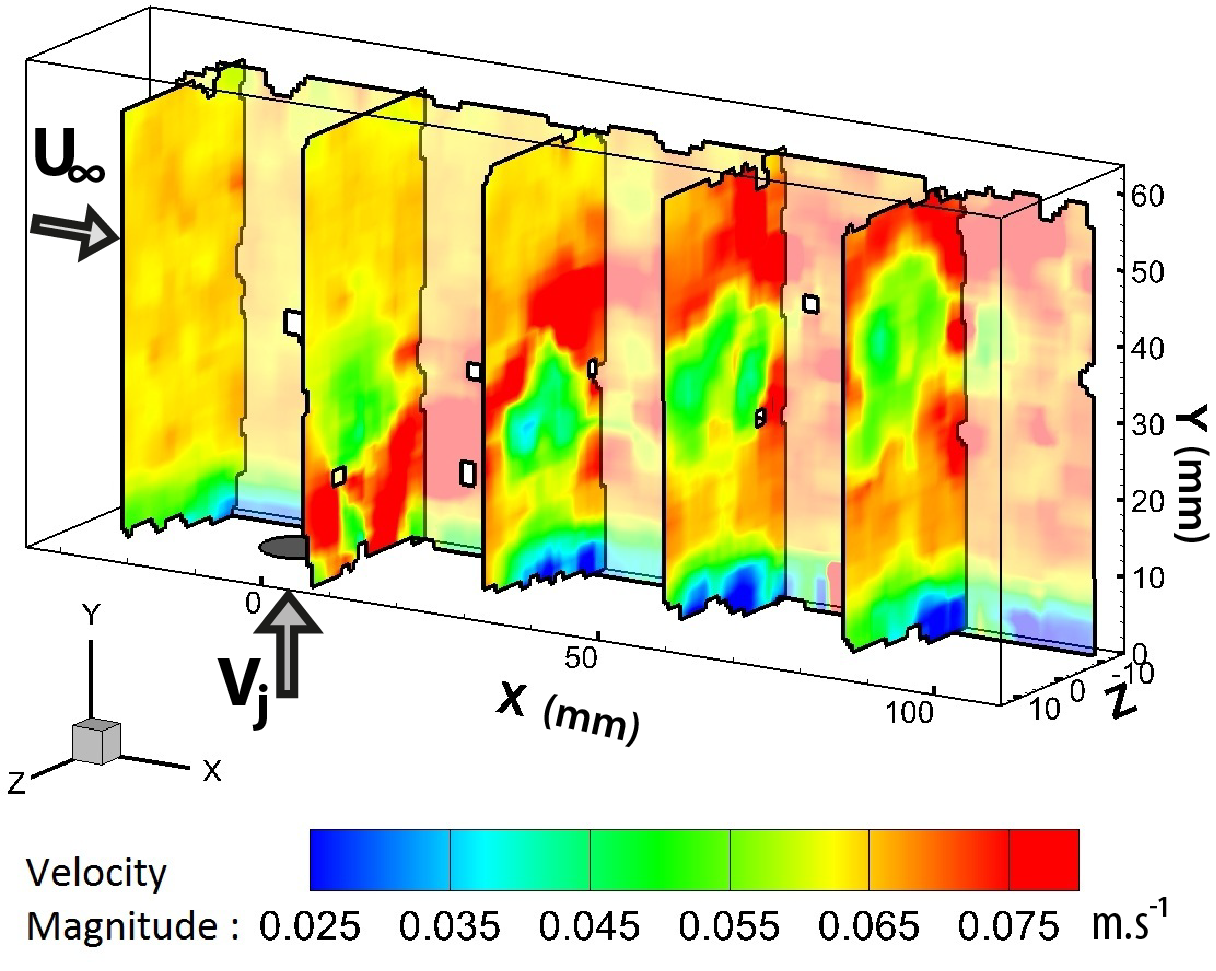}&
\includegraphics[width=0.5\textwidth]{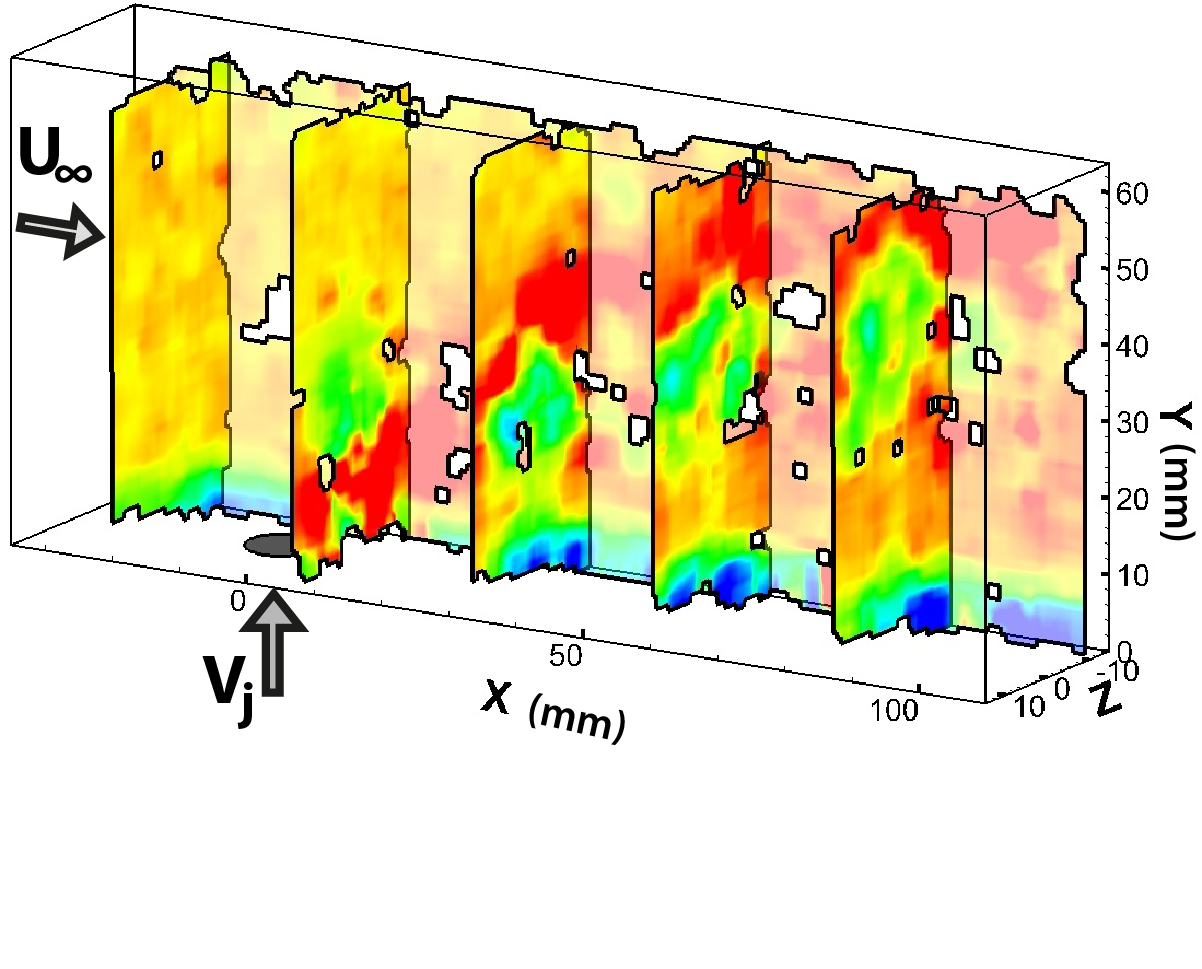}\\
a)&b)
\end{tabular}
\end{center} 
\caption{Jet in crossflow experiment. Transversal plane colored in velocity magnitude located at $z=-1.5$d and longitudinal planes colored in velocity magnitude located at $X=-3\mathrm{d_{jet}}$, 1$\mathrm{d_{jet}}$, 5$\mathrm{d_{jet}}$, 9$\mathrm{d_{jet}}$ et 13$\mathrm{d_{jet}}$.  The same instantaneous velocity field is interpolated in order to get a final resolution of : a) 1 vector every 1 mm. b) 1 vector every 0.9 mm.}\label{Fig19}
\end{figure}

The Fig. \ref{Fig19} presents two  velocity fields interpolated from the same instantaneous raw velocity field. In the Fig. \ref{Fig19}a  the spatial resolution of the velocity field is 0.1 cm whereas it is 0.09 cm for the Fig. \ref{Fig19}b. 
The edges of the data are underlined using thick black lines to better visualize the empty voxels regions (where the interpolation have not be carried out). 
The visualization slice is located near the back-side of the measurement volume, the furthest from the cameras.
One notice on Fig. \ref{Fig19}a that the velocity field is full, except a few voxels. The covering percentage for this field is of 98.5$\%$ with 16 861 raw velocity vectors obtained from images with 51 064 2D detected particles ($Cp_{2D} = 682.2$ part.cm$^{-2}$). This leads to an efficiency of the whole processing $R_{Eff\ Proc} =16861/51064 \approx 1/3$, which is a clear improvements compared to previous studies ($1/4$ for \citet{Sharp2010} and $1/5$ for \citet{Wolf2011}). This value is slightly lower than the expected 99$\%$. It is not surprising considering that the study of section 4 was carried out under optimal conditions: uniform flow, no velocity gradient, $\delta t$ between two images adjusted to have an optimal pixel displacement $\delta p$.

On the other hand, many features of the jet in crossflow make the tracking of the particles between two frames much more difficult and complex. The jet itself as well as the many vortex systems create very sudden velocity changes in the fluid and the multiple shear layers impose a wide range of velocity gradients. All that should lower the covering percentage. Even if we observe such a reduction, it is very weak. This result illustrates the robustness of the relaxation method algorithms.\\

The Fig. \ref{Fig19}b corresponds to the same instantaneous velocity field, but interpolated with a slightly lower vector resolution (1 interpolated vector every 0.09 cm). Large holes corresponding to empty voxels are clearly visible. The covering percentage is  97.1$\%$: the whole measurement domain is no longer adequately interpolated.  This 1.4$\%$ difference is due to local inhomogeneities in the raw vector concentration $\mathrm{Cv_{3D}}$. Inside some voxels, the local raw vector concentration is not high enough to successfully interpolate a velocity vector in it.  

The measurements have been carried out properly because of the choice of $\mathrm{Cp_{2D}}=680\ part.cm^{-2}$. The unsuccessfully interpolated voxels are neither due to the screening phenomenon nor the loss of performance of the algorithms. 
Even if  there are no algorithmic decay nor optical screenings in this measurement, the particles in fluids layers closer to cameras are always slightly easier to detect and track than those farther from the cameras. This effect is seen in Fig. \ref{Fig10}. It leads to a small linear dependency between the local $\mathrm{Cp_{3D}}$ and the distance along the optical axis. It is then logical that the degradation of the interpolation coverage starts at the back of the measurement volume, as can be seen in Fig. \ref{Fig19}b.

This method together with Eq.  \ref{equationcouvertureCv3D} enable us to determine the minimum voxel size leading to a local $\mathrm{Cv_{3D}}$ of the raw velocity fields that makes the uniform interpolation possible with respect to the desired covering percentage. 
The poorer  interpolation in the back of the measurement volume (and related sudden drop of the covering percentage) is an important result. It shows that the empirical law (Eq.  \ref{equationcouvertureCv3D}) is still valid for complex flow and can still be used to determine the critical value of interpolation size $l_{vox}$ .

\section{Discussions and conclusion}

Volumetric velocimetry becoming more and more popular, the objective of this study was to propose a methodology optimizing the seeding concentration of tracer particles to get the best instantaneous (and averaged) velocity fields while avoiding optical screening. The influence of several physical parameters on the efficiency of volumetric measurements using the V3V system have been investigated. Even if  the experimental setup and related parameters obviously have an influence on the results, we demonstrate that it is possible through the optimization of the seeding to improve the spatial resolution of instantaneous volumetric measurements. This result is true for any other setup. If an experimentalist wants to use a given volumetric velocimetry system, he can still follow the guidelines given in this paper. He will know it is possible to visualize instantaneous  3D vortical structures with a good spatial resolution and he will know how to do it.\\

Once the setup has been optimized (homogenization of the light intensity inside the measurement volume, careful alignment of the measurement volume with the calibration volume, etc), the concentration of particles becomes the most critical parameter to optimize volumetric velocimetry. In addition to the real concentration in particles $\mathrm{C_{m}}$ several other concentrations have been introduced to measure the number of particles identified during each processing steps.

When increasing the concentration of particles in the flow, for a given measurement volume, the first phenomenon  is the algorithmic decay which lowers the performances of the tracking algorithm. Larger particle concentrations lead to the setting of optical screening and intensity homogenization of the background making difficult the particles identification and, as a consequence, their tracking. It should be noted that it should also be avoided at all cost for mean velocity fields measurements. Even if a large number of instantaneous fields leads to a final mean field without \lq\lq{}holes\rq\rq{}, optical screening introduces an inhomogeneity in the spatial repartition of 3D particles and raw velocity vector field. In this case, the convergence of the statistical properties of the velocity field in the volume may also become inhomogeneous.

The optimal concentration leading to the highest resolution has been sought. We found that for $620<\mathrm{Cp_{2D}}<840$ particles.cm$^{-2}$, the $\mathrm{Cv_{3D}}$ is optimal for our measurements. Taking that into account, it leads to an absolute concentration $3.8\times 10^{-2}<\mathrm{Cp_{2D}}<5.2\times 10^{-2}\ ppp$. Since the whole experimental setup has been carefully optimized accordingly to the software and hardware constraints of the V3V system  (camera positioning, calibration, particle size), even if these 2D concentration values are relative to our experiments, they also correspond to an optimal choice for V3V measurements.

A relation between the covering percentage of interpolation and the adimensionalized $\mathrm{Cv_{3D}}$ was also found. Using this relation, it becomes possible, for a given optimal $\mathrm{Cv_{3D}}$, to estimate in return the size of the interpolation mesh corresponding to the required covering percentage. For the optimal $\mathrm{Cv_{3D}}$ this relation allows either to determine the smallest interpolation voxel size which provides the desired covering percentage for a given illumination thickness, or to determine the largest illumination thickness which provides a desired covering percentage for a given voxel size. Equation \ref{equationcouvertureCv3D} corresponding to very simple configuration, it is expected that the covering percentage will be lower for more complex flows. 

This methodology is finally applied to a complex test case, a jet in crossflow. This complex 3D flow brings into play many shear-layers together with multiple swirling structures. The topology of the mean velocity field described in many other experimental and numerical studies is recovered. Very good instantaneous 3D velocity fields are also measured. 

It was also shown that the calculated interpolation mesh size corresponds  to a critical mesh size below which the covering percentage of the interpolated velocity fields starts to decrease. The obtained covering percentage (98.5$\%$) is very close from the expected value (99$\%$). It clearly demonstrates that the tracking algorithms remain efficient in the range of parameters selected and are well adapted to complex flows. 

Finally, optimization of the concentration of the particles is a key point to improve the ratios of processing efficiency. Very good ratios have been obtained (between $1/3$ and $1/2$), much better than the ones obtained in previous studies which were ranging between 1/4 \citep{Sharp2010} and 1/5 \citep{Wolf2011}. It illustrates the importance of optimizing the setup and particles seeding to obtain good instantaneous volumetric three-components velocity fields.

\bibliographystyle{elsarticle-num-names}

\bibliography{BiblioOptMethodo4}
\end{document}